\documentclass[aps,prl,twocolumn,showpacs,psfig,superscriptaddress,longbibliography]{revtex4-1}

\usepackage{times}
\usepackage{graphicx}
\usepackage{float}
\usepackage{latexsym,amsmath,amssymb,bm,euscript}
\usepackage{color}
\usepackage{subfigure}
\usepackage{epstopdf}
\usepackage[colorlinks=true,linkcolor=blue,citecolor=blue]{hyperref}
\usepackage{hyperref}
\usepackage{soul}
\usepackage[normalem]{ulem}
\usepackage{mathrsfs}
\usepackage{amsmath}
\usepackage{xspace}

\newcommand{\citex}[1]{%
\begin{NoHyper}\citeauthor{#1} (\citeyear{#1})\,\end{NoHyper}%
\cite{#1}}

\newcommand{\Eq}[1]{Eq.~\eqref{#1}}

\newcommand{\Fig}[1]{Fig.~\ref{#1}}
\newcommand{\Figs}[1]{Figs.~\ref{#1}}

\def\Sq{\ensuremath{S(q)}\xspace} 
\def\Dstar{\ensuremath{D^\ast}\xspace}
\def\Th{\ensuremath{T_h}\xspace}
\def\Tl{\ensuremath{T_l}\xspace}

\def\Jtc{\ensuremath{J_t^c}\xspace}

\def\Tchi{\ensuremath{T\!\chi_0}\xspace}

\newcommand{\sparagraph}[1]{\subsubsection{#1}}

\graphicspath{{./Figs/}}

\begin{document}

\title{Two-temperature scales in the triangular-lattice Heisenberg antiferromagnet}

\author{Lei Chen}
\affiliation{Department of Physics, Key Laboratory of Micro-Nano Measurement-Manipulation and Physics (Ministry of Education), Beihang University, Beijing 100191, China}

\author{Dai-Wei Qu}
\affiliation{Department of Physics, Key Laboratory of Micro-Nano Measurement-Manipulation and Physics (Ministry of Education), Beihang University, Beijing 100191, China}

\author{Han Li}
\affiliation{Department of Physics, Key Laboratory of Micro-Nano Measurement-Manipulation and Physics (Ministry of Education), Beihang University, Beijing 100191, China}

\author{Bin-Bin Chen}
\affiliation{Department of Physics, Key Laboratory of Micro-Nano Measurement-Manipulation and Physics (Ministry of Education), Beihang University, Beijing 100191, China}
\affiliation{{Munich Center for Quantum Science and Technology (MCQST),
Arnold Sommerfeld Center for Theoretical Physics (ASC) and
Center for NanoScience (CeNS), Ludwig-Maximilians-Universit\"at M\"unchen,
Fakult\"at f\"ur Physik, D-80333 M\"unchen, Germany.}}

\author{Shou-Shu Gong}
\affiliation{Department of Physics, Key Laboratory of Micro-Nano Measurement-Manipulation and Physics (Ministry of Education), Beihang University, Beijing 100191, China}

\author{Jan von Delft}
\affiliation{{Munich Center for Quantum Science and Technology (MCQST),
Arnold Sommerfeld Center for Theoretical Physics (ASC) and
Center for NanoScience (CeNS), Ludwig-Maximilians-Universit\"at M\"unchen,
Fakult\"at f\"ur Physik, D-80333 M\"unchen, Germany.}}

\author{Andreas Weichselbaum}
\email{weichselbaum@bnl.gov}
\affiliation{Department of Condensed Matter Physics and Materials
Science, Brookhaven National Laboratory, Upton, New York 11973-5000, USA}
\affiliation{{Munich Center for Quantum Science and Technology (MCQST),
Arnold Sommerfeld Center for Theoretical Physics (ASC) and
Center for NanoScience (CeNS), Ludwig-Maximilians-Universit\"at M\"unchen,
Fakult\"at f\"ur Physik, D-80333 M\"unchen, Germany.}}

\author{Wei Li}
\email{w.li@buaa.edu.cn}
\affiliation{Department of Physics, Key Laboratory of Micro-Nano Measurement-Manipulation and Physics (Ministry of Education), Beihang University, Beijing 100191, China}
\affiliation{International Research Institute of Multidisciplinary Science, Beihang University, Beijing 100191, China}

\begin{abstract}
The anomalous thermodynamic properties
of the paradigmatic frustrated
spin-1/2 triangular lattice Heisenberg 
antiferromagnet (TLH) has remained 
an open topic of research over decades,
both experimentally and theoretically. Here we further
the theoretical understanding based on 
the recently developed, powerful
exponential tensor renormalization group (XTRG) method
on cylinders and stripes in 
a quasi one-dimensional (1D) setup, 
as well as a tensor product
operator approach directly in 2D.
The observed thermal properties of the TLH 
are in excellent agreement 
with two recent experimental measurements 
on the virtually
ideal TLH material Ba$_8$CoNb$_6$O$_{24}$.
Remarkably, our numerical simulations reveal 
two crossover temperature scales,
at $\Tl/J \sim 0.20$
and $\Th/J\sim 0.55$, with $J$ the 
Heisenberg exchange
coupling, which are also confirmed by 
a more careful inspection of the experimental 
data. We propose that 
in the intermediate regime between
the low-temperature scale $\Tl$ and the higher one \Th, 
the ``rotonlike" excitations
are activated with a strong chiral component
and a large contribution to thermal entropies.
Bearing remarkable resemblance to the renowned 
roton thermodynamics in liquid helium, 
these gapped excitations suppress the incipient 120$^\circ$ order
that emerges for temperatures below \Tl.
\end{abstract}
\date{\today}

\maketitle
\textit{Introduction.}
The triangular lattice Heisenberg (TLH) model is arguably
the most simple prototype of a frustrated quantum spin system.
It has attracted wide attention since Anderson's famous proposal
of a resonating valence bond (RVB) spin liquid state
\cite{Anderson1973}. The competition between RVB liquid versus
semiclassical N\'eel solid states raised great interest.
After decades of research, it is now widely accepted
that the TLH has noncollinear 120$^{\circ}$ order at $T=0$,
with a spontaneous magnetization \cite{Bernu1992a},
$m\simeq0.205$ \cite{Capriotti1999,White07}.
Nevertheless, the TLH has long been
noticed to possess \textit{anomalous}
thermodynamic properties \cite{Elstner-1993},
in the sense that thermal states down to
rather low-temperature regimes
behave more as a system with no indication of
an ordered ground state \cite{Elstner1994, Kulagin2013}.

Bipartite-lattice Heisenberg antiferromagnets (AFs)
such as the square-lattice Heisenberg (SLH) model, develop 
a semiclassical magnetic order at $T=0$ which is 
``melted" at any finite temperature 
according to the Mermin-Wagner theorem \cite{Mermin-Wagner}.
Nevertheless, the groundstate N\'eel order strongly influences low-temperature
thermodynamics in the so-called renormalized classical (RC) regime \cite{Chakravarty1988,Chakravarty1989},
where the spin-spin correlation length $\xi$
increases exponentially as $T$ decreases \cite{Beard1998,Kim1998,PhysRevLett.75.938,Greven1994}.

\begin{figure}[!tbp]
\includegraphics[angle=0,width=1\linewidth]{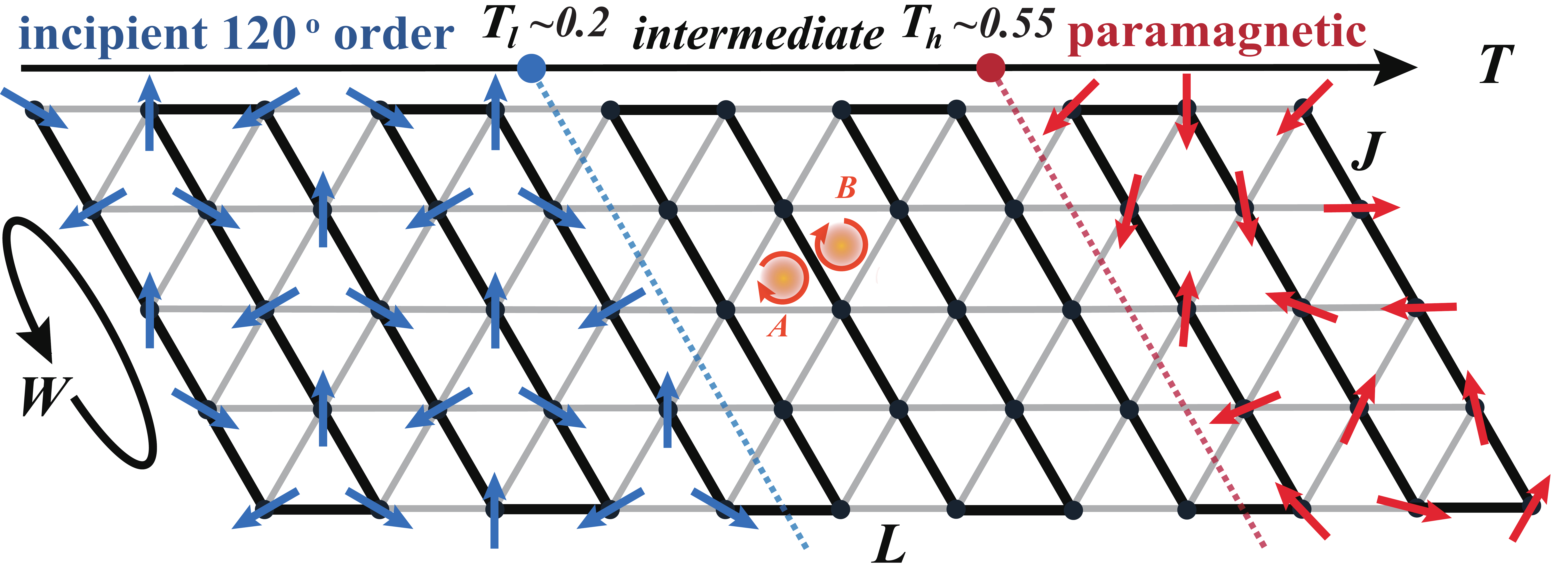}
\caption{(Color online) Uniform TLH with
nearest-neighbor (NN) coupling $J{=}1$
(which thus sets the unit of energy)
and lattice spacing $a{=}1$, with three schematically
depicted distinct regimes, separated by two cross-over
temperature scales, \Tl and \Th:
an incipient 120$^\circ$ ordered regime for $T<\Tl$ (left),
a paramagnetic regime for $T>\Th$ (right),
an intermediate regime 
(center), which is explored in detail in this paper.
The thick black line indicates the 1D snake order
adopted in the MPO-based XTRG.
When the system is wrapped into a cylinder
along the tilted left arrow, this is
referred to as YC geometry.
The clockwise oriented
circles in the center of the system
indicate chiral operators,
$\chi \equiv 2^3 \cdot S_a\cdot (S_b\times S_c)$,
acting on the enclosing triangle of sites $(a,b,c)$
in the order of the arrows,
as used for the calculation
of chiral correlations between 
the triangle pair A-B.
}
\label{Fig:PhDiag}
\end{figure}

In contrast, the thermodynamics of the TLH
strikingly differs in many
respects from that of SLH.
Based on high-temperature series expansion
(HTSE) results, both models show $c_V$
peaks at similar temperatures,
$\Th\simeq0.55$ (TLH) and $T_s\simeq0.6$ (SLH).
The SLH enters the RC regime for $T\lesssim T_s$
\cite{Beard1998,Kim1998}, whereas the TLH shows no
signature for incipient order and possesses
anomalously large entropies
at temperatures below \Th \cite{Elstner1994}.
  
The classical SLH and TLH models have a similar spin stiffness
$\rho_s$, and thus a similar constant, $C_\xi{\sim}\rho_s$,
in the correlation length, $\xi{\sim}\exp{(\frac{C_{\xi}}{T})}$,
as well as in the static structure factor at the ordering
wave vector, $S(K){\sim}\exp{(\frac{2C_{\xi}}{T})}$, with
$C_{\xi}{=}2 \pi \rho_s{=}1.571$ (SLH) \cite{Singh1989} and
$C_{\xi}{=}4 \pi \rho_s{=}1.748$ (TLH) \cite{Elstner-1993,Zheng-2006,PhysRevLett.68.1762}
in units of spin coupling $J$. 
However, the constant $C_{\xi}$ is significantly renormalized
by quantum fluctuations. 
For the SLH, the constant is reduced by about 30\%
to $C_{\xi}{\sim}1.13$,
while in the TLH it is reduced by an order of magnitude
down to $C_{\xi}{\sim}0.1$
\cite{Elstner-1993,Elstner1994}. 
The energy scale $E_\mathrm{RC}\equiv 2C_{\xi}$
naturally represents the onset of RC behavior and thus
incipient order.
Recent sign-blessing bold diagrammatic 
Monte Carlo (BDMC) simulations still show 
that the thermal states down to the lowest accessible
temperatures $T=0.375$ 
``extrapolate" to a disordered ground state
via a quantum-to-classical {correspondence} \cite{Kulagin2013}. 
%

Here, we exploit two renormalization group (RG) techniques
based on thermal tensor network states (TNSs) \cite{Li.w+:2011:LTRG,Chen.b+:2017:SETTN,Chen2018}: the exponential tensor RG (XTRG)
which we recently introduced 
based on one-dimensional (1D) matrix product operators (MPOs)  \cite{Chen2018},
and a tensor product operator (TPO)
approach \cite{Li.w+:2011:LTRG}.
XTRG is employed to simulate the
TLH down to temperatures $T<0.1$
on YC\,$W(\times L)$ geometries (see Fig.~\ref{Fig:PhDiag})
up to width $W=6$ with default $L=2W$,
and open strips [OS\,$W(\times L)$] with
fully open boundary conditions (OBCs) 
and default $L=W$ \cite{Supplementary}.

\textit{TLH thermodynamics.}
In \Fig{Fig:Thermos} we 
present our thermodynamical results from XTRG on
cylinder (YC) and open geometries (OS), as defined earlier.
In Fig.~\ref{Fig:Thermos}(a), we observe from YC5, OS6, 
and YC6 data that,  besides a high temperature 
round peak at $\Th \sim 0.55$, 
our YC data exhibit another peak 
(shoulder for OS6) at $\Tl \sim 0.2$. 
On YCs, the peak position \Tl stays
nearly the same when increasing $W$ from 5 to 6,
also consistent with the shoulder in OS6 
as well as in the experimental data. 
At the same time,
the low-temperature peak becomes slightly
weakened, yet towards the experimental data.
When compared to the two virtually 
coinciding experimental data sets,
YC6, TPO, earlier HTSE \cite{Elstner-1993},
and latest Pad\'e [6,6] data \cite{Rawl2017} 
all agree well for $T\gtrsim \Th$ and reproduce 
the round peak of $c_V$ at $\Th$.

The remarkable agreement of finite-size XTRG
with  experimental measurements can be ascribed 
to a short correlation length $\xi  
\lesssim 1$ lattice spacing for 
$T\gtrsim 0.4$ \cite{Supplementary}. 
Deviations from experiment
only take place below $\Tl$, 
suggesting significant finite-size effects 
due to larger $\xi$ in that regime.
Moreover, we have checked the dependence of
$\Tl$ on the cylinder length $L$ for YC6, and find that the
lower peak even gets slightly enhanced as $L$ 
increases.
In addition to YC and OS geometries, 
simulations on X cylinders also lead to the same
scenario  \cite{Supplementary}.



\begin{figure}[!tbp]
\includegraphics[angle=0,width=0.9\linewidth]{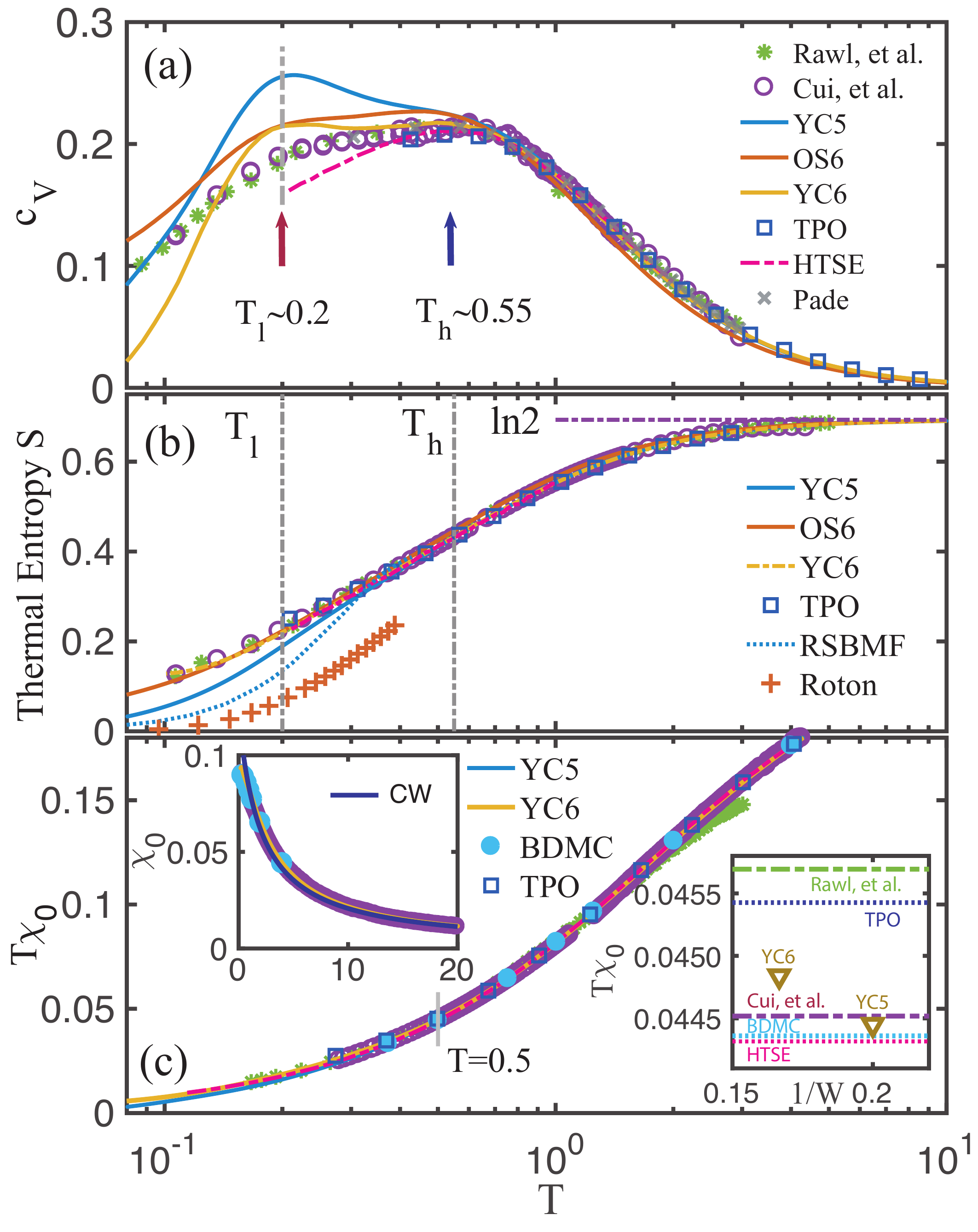}
\caption{(Color online)
    Simulated thermodynamics in comparison to experimental
    measurements, \citex{Cui2018} and \citex{Rawl2017},
    as well as earlier numerical results.
    The YC and OS data are obtained via XTRG by retaining up
    to $\Dstar=1000$ multiplets [$D\sim 4000$ U(1) states],
    and by a TPO method \cite{Supplementary} on infinite lattices,
     keeping up to 40 bond states.
(a) Specific heat, $c_V$, results benchmarked against HTSE
    \cite{Elstner-1993,Rawl2017} and experimental curves.
(b) The thermal entropy $S$ vs $T$, together with the
    reconstructed Schwinger boson mean field (RSBMF)
    \cite{Mezio2012}, and ``roton" contributions \cite{Zheng-2006}.
(c) Uniform magnetic susceptibility \Tchi vs $T$,
    shown with BDMC data \cite{Kulagin2013}. The left top inset
    compares $\chi_0$ to Curie-Weiss (CW)
    $\chi_0 = C/(T+\theta)$ in a wide temperature range,
    where $C=1/4$ and 
    $\theta=2.06$. In the right bottom inset we further
    compare various \Tchi values at $T=0.5$. The
    magnetic moment per Co is assumed $\simeq 2 \mu_B$, with
    Land\'e factor $g\simeq4.13$ \cite{Cui2018}.
}
\label{Fig:Thermos}
\end{figure}

In Fig.~\ref{Fig:Thermos}(b), we present
our data on thermal entropy, again directly juxtaposed
with experimental as well as 
previous theoretical results. 
Whereas the YC5 data deviate at 
$T\lesssim0.3$ due to finite-size
effects, we observe good
agreement between the two experimental 
data sets with our TPO
results down to $T_l$, and with
$W=6$ data (OS6 and YC6)
down to the lowest temperatures 
in the measurements. 
Notably, the thermal entropy per site $S$
is about 1/3 of the high-$T$ limit,  
$S_\infty=\ln{2}$, at temperatures as low as $T\simeq 0.2$
where, for comparison, for SLH $S$ is almost 
zero at the same temperature \cite{Elstner1994}. 
We emphasize 
that \Fig{Fig:Thermos}(b) is a direct 
comparison without any fitting, since 
the only parameter $J$ has also been 
determined and thus fixed as 1.66\,K 
in the experiments \cite{Rawl2017,Cui2018}. 
Nevertheless, since the 
experimental data of $S$ are determined by 
integrating $c_V/T$, starting from the 
lowest accessible temperature $T_x$,
systematic vertical shifts for the curves from 
Refs.~\cite{Rawl2017} and \cite{Cui2018}
are necessary to reach the known large-$T$ limits.
This results in residual entropies of
$S(T_x)=0.045$ and $0.06$ at temperatures
$T_x=0.06$ and $0.08$\,K, 
for Refs.~\cite{Rawl2017} and \cite{Cui2018}, respectively.  
Note that the large entropy due to quantum frustration 
at low $T$ is not properly described 
in previous theories, e.g., 
RSBMF \cite{Mezio2011, Mezio2012} 
as shown in \Fig{Fig:Thermos}(b). 

Figure~\ref{Fig:Thermos}(c) presents our results for the average
magnetic susceptibility. 
Both data sets, YC5 and YC6, agree quantitatively with the
experimental results, as well as HTSE data \cite{Elstner-1993},
from high temperatures down to 
$T \lesssim 0.1$, well beyond 
state-of-the-art BDMC results that reach 
down to $T=0.375$ \cite{Kulagin2013}. 
In the left top inset of Fig.~\ref{Fig:Thermos}(c), 
we also include a Curie-Weiss (CW)
fit for $T\gtrsim1$, 
resulting in the positive Weiss constant 
$\theta \approx 2 J$. 
In the right bottom inset, we compare the 
\Tchi value at $T=0.5$,
and find the various numerical and
experimental results all agree, 
up to three significant digits. 

\textit{Two-temperature scales.}
As schematically depicted in Fig.~\ref{Fig:PhDiag},
we uncover a two-temperature-scale scenario in the TLH.
This confirms that the 120$^\circ$ order plus 
magnon excitations is not sufficient to describe
TLH thermodynamics. 
References.~\cite{Zheng2006PRL,Starykh2006}
argued that the TLH also has additional 
types of excitations which are gapped, 
with the minimum of their quadratic 
dispersion at finite momentum,
and referred to these as ``rotonlike excitations"  (RLEs), 
since their dispersions are reminiscent of that 
known for vortexlike excitations in He$^4$ \cite{Feynman1954}.
Excitations with this type of dispersion have 
recently also been observed in neutron 
scattering experiments of TLH materials \cite{Ma2016,Ito2017}. 
RLEs evidently play an important role in the 
intermediate-temperature regime
in Fig.~\ref{Fig:PhDiag},
but their precise nature has not yet been fully elucidated.

RLEs, although missed in the linear spin-wave theory, can be
well captured by including $1/S$ corrections in calculating the 
magnon dispersions \cite{Starykh2006,Chernyshev-2016,
Chernyshev-2009} and dynamical correlations \cite{Mourigal-2013,Luo-2015}.
Other proposals have also
been put forward to understand RLEs, 
including the vortex-antivortex excitation \cite{Alicea06}
with signatures already in the classical TLH phase diagram
versus finite temperature \cite{Kawamura10,Gvozdikova11,Seabra11,Popov2017},
(nearly deconfined) spinon-antispinon pair \cite{Zheng2006PRL,Zheng-2006,Piazza2014}, 
and magnon-interaction-stabilised excitations \cite{Mourigal-2013, Verresen-2018,Verresen2018b}. 

First, the RLE quadratic band with 
a finite gap $\Delta{\sim} 0.55\,J$ contributes to a 
very prominent peak in the density of states around
$\Delta$ \cite{Zheng-2006}. 
This coincides with the high-temperature scale
$\Th \sim \Delta$ here. Therefore a
possible connection of RLEs 
to the thermodynamic anomaly in TLH 
has been suggested earlier
\cite{Zheng-2006,Chernyshev-2009}.
Second, the RLEs themselves 
only start to significantly contribute to the entropy
above $\Tl$ [`Roton' entry in \Fig{Fig:Thermos}(b),
with data taken from Ref.~\cite{Zheng-2006}].
This suggests that the RLEs are activated 
in the intermediate temperature regime, i.e., 
$\Tl\lesssim T\lesssim\Th$.
Consequently, the onset
of incipient magnetic order
is postponed to a clearly lower temperature 
$T_l\sim0.2$, 
which is remarkably close to
previous HTSE studies, 
where $E_\mathrm{RC}\sim 0.2\,J$
sets the energy scale of classical correlation
\cite{Elstner-1993} as discussed earlier.

\begin{figure}[!tbp]
\includegraphics[angle=0,width=1\linewidth]{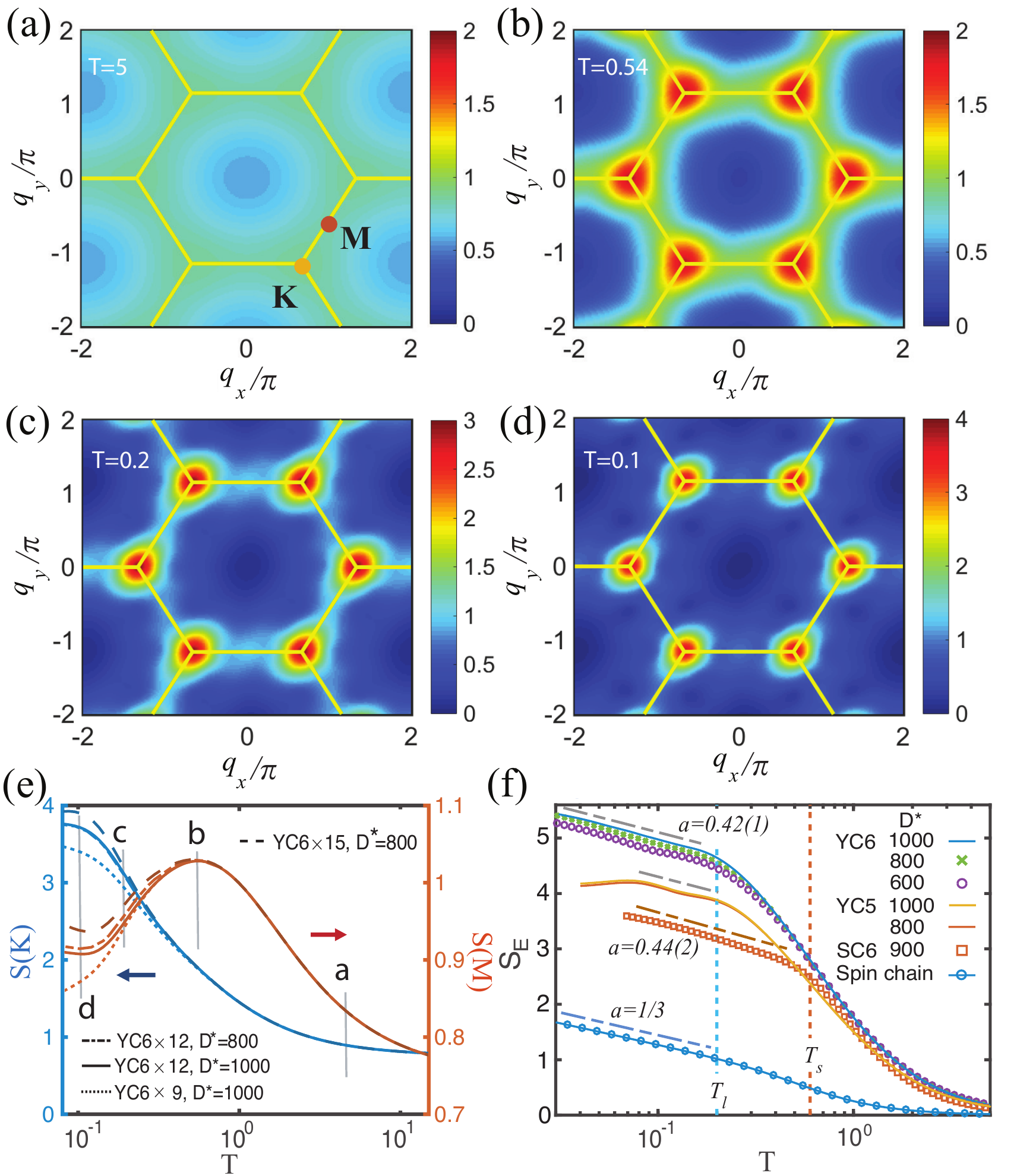}
\caption{(Color online)
(a-d) Structure factor on YC$6\times12$ lattice,
i.e., with $q_y$ pointing along the direction of the cylinder,
at temperatures $T=5$, $0.54$, $0.2$, and $0.1$,
respectively, [vertical gray lines in (e)]. 
(e) $\Sq$ vs $T$ at momenta
    $q=K$ and $M$ where the legend holds for both data sets.
(f) $S_E$ vs $T$, where the tilted dashed lines indicate
    the logarithmic scaling $S_E = a \ln(\beta) + b$, where
    the slopes $a$ seen for the TLH are similar to that for
    the SLH (SC6 data). The vertical dashed line 
    labels the low-temperature scale
    $\Tl\sim0.2$ for TLH and the only temperature scale
    $T_s\sim0.6$ for SLH.  SC$6\times12$ stands for a
    $W=6,L=12$ square cylinder, and $S_E$ scaling in the
    Heisenberg chain (length $L=200$) is also plotted as a
    comparison.
}
\label{Fig:SF}
\end{figure}

\textit{Spin structure factors.}
In order to shed light into the spin configurations across
the intermediate regime, we turn to the
temperature dependent static structure factor,
$S(q) \equiv \sum_{j} e^{-i q \cdot r_{0j}}
\, \langle {\mathbf{S}}_{0} \cdot \bold{S}_j \rangle_T$
where $r_{0j} \equiv r_j - r_0$ with $r_j$ the lattice
location of site $j$, and $S(q)\in \mathbb{R}$ due to
lattice inversion symmetry.
%
There are two further high-symmetry points of interest, $q=K$ and $M$,
as marked in \Fig{Fig:SF}(a). Up to symmetric reflections,
$K{\equiv}(\frac{2\pi}{3}, \frac{2\pi}{\sqrt{3}})$ relates
to $120^{\circ}$ non-collinear order, whereas $M{\equiv}(0,
\frac{2\pi}{\sqrt{3}})$ relates to nearest-neighbor (stripe)
AF correlations. The latter have also been related to
RLEs which feature band minima at the $M$ points
\cite{Zheng2006PRL,Starykh2006,Ghioldi2018}.

In \Figs{Fig:SF}(a-d) we show
the overall landscape of \Sq. 
With decreasing temperature, 
$\Sq$ changes from rather
featureless in \Fig{Fig:SF}(a),
to showing bright regions in the vicinity of the six
equivalent $K$ points
as well as enhanced intensity at the $M$ points
at $T\sim\Th$ in \Fig{Fig:SF}(b).
Even at $T\sim \Tl$ in \Fig{Fig:SF}(c), 
one can still recognize an
enhanced intensity $S(M)$,
which fades out
eventually when $T$ is decreased below $\Tl$
in \Fig{Fig:SF}(d).
A quantitative comparison is
given in \Fig{Fig:SF}(e).

From \Fig{Fig:SF}(e), we observe that
$S(K)$ increases mono\-tonously as $T$ decreases.
It is featureless around \Th, and eventually saturates
at the lowest $T$ due to finite system size.
For $T>T_l$, $S(K)$ increases only
slowly with decreasing temperature, and is independent
of length $L$. It therefore shows no 
signature of incipient order there. 
For $T<T_l$, $S(K)$ rapidly increases, 
which eventually saturates
with decreasing $T$ in an $L$-dependent manner, 
due to finite-size effects.
%

Furthermore, we observe from \Fig{Fig:SF}(e)
that $S(M)$ develops a well-pronounced maximum around \Th.
The maximum is already stable
with system size, hence can be considered a feature in the
thermodynamic limit. This is consistent with a
picture that RLEs are activated
near the $M$ points. 

\textit{MPO entanglement.}
The two-energy-scale scenario also leaves a
characteristic trace in
the entanglement entropy $S_E$,
computed at a bond (near the center) 
of the MPO
\cite{Verstraete04purif,Feiguin05purif,Chen2018}.
Gapless low-energy excitations in 1+1D conformal field theory (CFT)
can give rise to a logarithmic increase of the entanglement,
$S_E \propto -\frac{c}{3} \ln{T}$ 
with $c$ the conformal central charge 
\cite{Barthel.t:2017:FiniteT,Dubail17,Chen2018}.
One can also observe logarithmic $S_E$ behavior
in the 2D SLH model, related to
the spontaneous SU(2) symmetry breaking (at $T=0$)
\cite{Chen2018}, as also added for
reference (``SC6" data) in \Fig{Fig:SF}(f).

We find similar behavior of 
the $S_E$ profiles of the TLH 
on YC5 and YC6 geometries
in \Fig{Fig:SF}(f) down to $T=0.04$,
with bond dimension $\Dstar
\lesssim 1000$ multiplets ($D\sim4\Dstar$ states).
Interestingly, the lower-energy scale
$\Tl \sim 0.2$ (vertical dashed line) signals the onset of
logarithmic entanglement scaling versus $T$,
which in agreement with \Fig{Fig:Thermos}(a)
already coincides for $W=5$ and $6$. For YC5, the
window with logarithmic entanglement is rather narrow,
below of which $S_E$ saturates as
we already approach the ground state.
For YC6, the entanglement
continues to increase down to our lowest temperature
$T=0.03$. We associate the logarithmic $S_E$ behavior with
the onset of incipient order, which is closely related to
SU(2) symmetry breaking 
at $T=0$ that gives rise, e.g., to a $1/(N=LW)$
level spacing in the low-energy
tower of states \cite{Bernu1992a}.
%
Concomitantly, we also observe a
qualitative change of behaviors
in the entanglement spectra
at $T_l$ \cite{Supplementary}.

\textit{Scalar chiral correlations.}
\begin{figure}[!tbp]
\includegraphics[angle=0,width=0.9\linewidth]{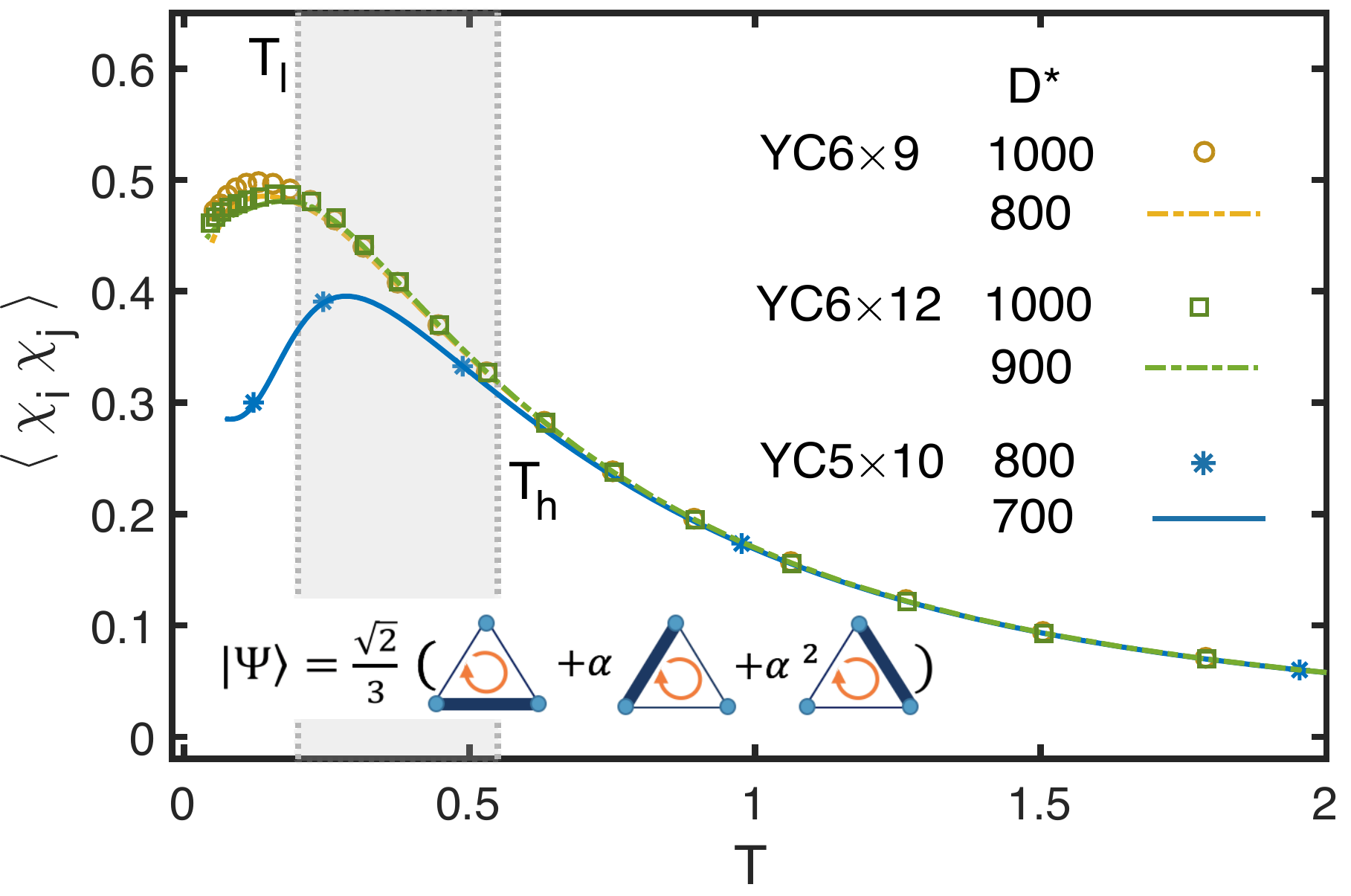}
\caption{(Color online)
Chiral correlations on cylinders, YC5 and YC6
(for YC4, see Ref.~\cite{Supplementary}).
The inset represents the
eigenstates $\Psi$ (and $\Psi^\ast$)
of the chiral operator $\chi$ (\Fig{Fig:PhDiag})
with non-zero eigenvalues $\pm \sqrt{12}$.
They have 
total spin $S=1/2$, and hence are
superpositions of configurations with two-site singlet dimers
(thick lines) whose signs are fixed in clockwise order (arrow).
Having $\alpha=\exp(2\pi i/3)$, this demonstrates the chiral
nature.
}
\label{Fig:Chiral}
\end{figure}
Chiral correlations in the TLH have raised great interest
since the proposal of a Kalmeyer-Laughlin
chiral spin liquid \cite{Kalmeyer1987}. 
Intriguingly, recent $T=0$ studies on the fermionic triangular
lattice Hubbard model proposed a chiral intermediate phase
versus Coulomb repulsion which thus breaks
time reversal symmetry 
\cite{Szasz2018}. While debated \cite{Shirakawa17},
we take this as a strong motivation to also study
traces of chiral correlations in the TLH at finite $T$.

In \Fig{Fig:Chiral}, we present the chiral correlation
$\langle \chi_i \chi_j \rangle$ between two
nearest-triangles $i,j$
in the system center, as defined with \Fig{Fig:PhDiag}.
This shows that chiral correlations are weak
in both high- and low-temperature limit, 
while they become strong \cite{Szasz2018}
in the intermediate temperature regime,
with a peak around \Tl. 
Below \Tl, the chiral correlations drop strongly, 
giving way to the buildup of coplanar incipient order. 


\textit{Discussion.} 
Our study suggests a tight connection between
RLEs and chiral correlations in the intermediate regime
$\Tl \lesssim T \lesssim \Th$ (cf. \Fig{Fig:Chiral}).
In this sense, we speculate that 
RLEs activated in the intermediate-temperature regime
indicate phase-coherent rotating dimers,
as schematically sketched with \Fig{Fig:Chiral}.
Given that the complex phase of the dimers ``rotates"
by $2\pi$, this suggests a possible link to a topological,
vortexlike nature of the RLEs. Moreover, it
resembles Feynman's notion of rotons in terms of quantized
vortices 
%
in He$^4$ \cite{Feynman1954}
via an 
exact mapping of TLH to a system of hardcore bosons. 
The latter further underlines the striking 
analogy between the anomalous thermodynamics of the TLH 
and the renowned roton thermodynamics in He$^4$
\cite{Kramers1952,Bendt1959}. 

The low-energy scale \Tl can be tuned by deforming
the Hamiltonian, e.g., by altering the level of frustration
by adding a next-nearest $J_2$ coupling to the TLH.
We see that increasing $J_2$ reduces \Tl,
as well as the height of the corresponding peak
in the specific heat, {suggesting that the RLE gap is 
decreasing and the influence can thus spread down 
to even lower-temperature/energy scales, in
consistency with dynamical studies 
of the $J_1$-$J_2$ TLH \cite{Ghioldi2015,Ferrari2019}.}
In addition, TLH can be continuously deformed
into the SLH, where \Tl increases and eventually merges
with \Th once sufficiently close to the SLH.
We refer more details to the Supplemental Materials \cite{Supplementary}.

\textit{Outlook.}
A detailed study of the microscopic nature of RLEs,
e.g., via dynamical correlations at finite temperature,
is beyond the scope of the present paper,
and is thus left for future research.
Further stimulating insights and possible
superfluid analogies are also expected
from an analysis of the interplay of external magnetic
fields and thermal fluctuations in TLH \cite{Griset-2011,Starykh-2015}
with clear experimental relevance 
\cite{Cui2018}.

\begin{acknowledgments}
\textit{Acknowledgments}. 
WL and LC would like to thank Yi Cui and Wei-Qiang Yu
for providing their original data of
Ba$_8$CoNb$_6$O$_{24}$.
WL is indebted to Lei Wang, Zi Cai, Hong-Hao Tu, 
Zheng-Xin Liu, Xue-Feng Zhang, Bruce Normand, and Jie Ma
for stimulating discussions.
This work was supported by the National
Natural Science Foundation of China
(Grant No. 11504014, 11834014, and 11874078)
and supported by the Deutsche Forschungsgemeinschaft (DFG, German Research Foundation) 
under Germany's Excellence Strategy -- EXC-2111 -- 390814868.
B.-B.C. was supported by the German Research foundation,
DFG WE4819/3-1. 
A.W. was funded by DFG WE4819/2-1 and
DOE DE-SC0012704 without temporal overlap. 
W.L. and S.-S.G. were supported by the 
Fundamental Research Funds for the Central Universities.
\end{acknowledgments}

%

\newpage\mbox{}\pagebreak

\setcounter{equation}{0}
\setcounter{figure}{0}
\setcounter{table}{0}
\setcounter{page}{1}

\makeatletter

\renewcommand{\theequation}{S\arabic{equation}}
\renewcommand{\thefigure}{S\arabic{figure}}
\renewcommand{\bibnumfmt}[1]{[#1]}
\renewcommand{\citenumfont}[1]{#1}
\widetext
\begin{center}
\textbf{\large Supplemental Materials: \\ Two-Temperature Scales in the Triangular-Lattice Heisenberg Antiferromagnet}

\author{Lei Chen}
\affiliation{Department of Physics, Key Laboratory of Micro-Nano Measurement-Manipulation and Physics (Ministry of Education), Beihang University, Beijing 100191, China}

\author{Dai-Wei Qu}
\affiliation{Department of Physics, Key Laboratory of Micro-Nano Measurement-Manipulation and Physics (Ministry of Education), Beihang University, Beijing 100191, China}

\author{Han Li}
\affiliation{Department of Physics, Key Laboratory of Micro-Nano Measurement-Manipulation and Physics (Ministry of Education), Beihang University, Beijing 100191, China}

\author{Bin-Bin Chen}
\affiliation{Department of Physics, Key Laboratory of Micro-Nano Measurement-Manipulation and Physics (Ministry of Education), Beihang University, Beijing 100191, China}
\affiliation{Munich Center for Quantum Science and Technology (MCQST),
Arnold Sommerfeld Center for Theoretical Physics (ASC) and
Center for NanoScience (CeNS), Ludwig-Maximilians-Universit\"at M\"unchen,
Fakult\"at f\"ur Physik, D-80333 M\"unchen, Germany.}

\author{Shou-Shu Gong}
\affiliation{Department of Physics, Key Laboratory of Micro-Nano Measurement-Manipulation and Physics (Ministry of Education), Beihang University, Beijing 100191, China}

\author{Jan von Delft}
\affiliation{Munich Center for Quantum Science and Technology (MCQST),
Arnold Sommerfeld Center for Theoretical Physics (ASC) and
Center for NanoScience (CeNS), Ludwig-Maximilians-Universit\"at M\"unchen,
Fakult\"at f\"ur Physik, D-80333 M\"unchen, Germany.}

\author{Andreas Weichselbaum}
\email{weichselbaum@bnl.gov}
\affiliation{Department of Condensed Matter Physics and Materials
Science, Brookhaven National Laboratory, Upton, NY 11973-5000, USA}
\affiliation{Munich Center for Quantum Science and Technology (MCQST),
Arnold Sommerfeld Center for Theoretical Physics (ASC) and
Center for NanoScience (CeNS), Ludwig-Maximilians-Universit\"at M\"unchen,
Fakult\"at f\"ur Physik, D-80333 M\"unchen, Germany.}

\author{Wei Li}
\email{w.li@buaa.edu.cn}
\affiliation{Department of Physics, Key Laboratory of Micro-Nano Measurement-Manipulation and Physics (Ministry of Education), Beihang University, Beijing 100191, China}
\affiliation{International Research Institute of Multidisciplinary Science, Beihang University, Beijing 100191, China}

\date{\today}

\maketitle

\end{center}

\setcounter{equation}{0}
\setcounter{figure}{0}
\setcounter{table}{0}
\setcounter{page}{1}

\makeatletter

\renewcommand{\theequation}{S\arabic{equation}}
\renewcommand{\thefigure}{S\arabic{figure}}
\renewcommand{\bibnumfmt}[1]{[#1]}
\renewcommand{\citenumfont}[1]{#1}

\section{Finite-Temperature Renormalization Group Approaches}
\label{Sup:TTN}

In this section, we recapitulate the two thermal tensor network
approaches employed in the present study, i.e, the exponential tensor
renormalization group (XTRG) based on matrix product operators
(MPOs) and the tensor product operator (TPO) approach. The
former is highly controllable while restricted to finite-size
systems; the latter can be applied directly in the thermodynamic
limit, but is constrained by finite bond dimensions as well as an
approximate optimization scheme. Nevertheless, we
exploit a combination of 1D/2D tensor network RG approaches to
study the triangular lattice Heisenberg (TLH) model,
and find the results of the two methods consistent.

\subsection{Exponential Tensor Renormalization Group:
A matrix product operator approach}

\begin{figure}[htbp]
\includegraphics[angle=0,width=0.55\linewidth]{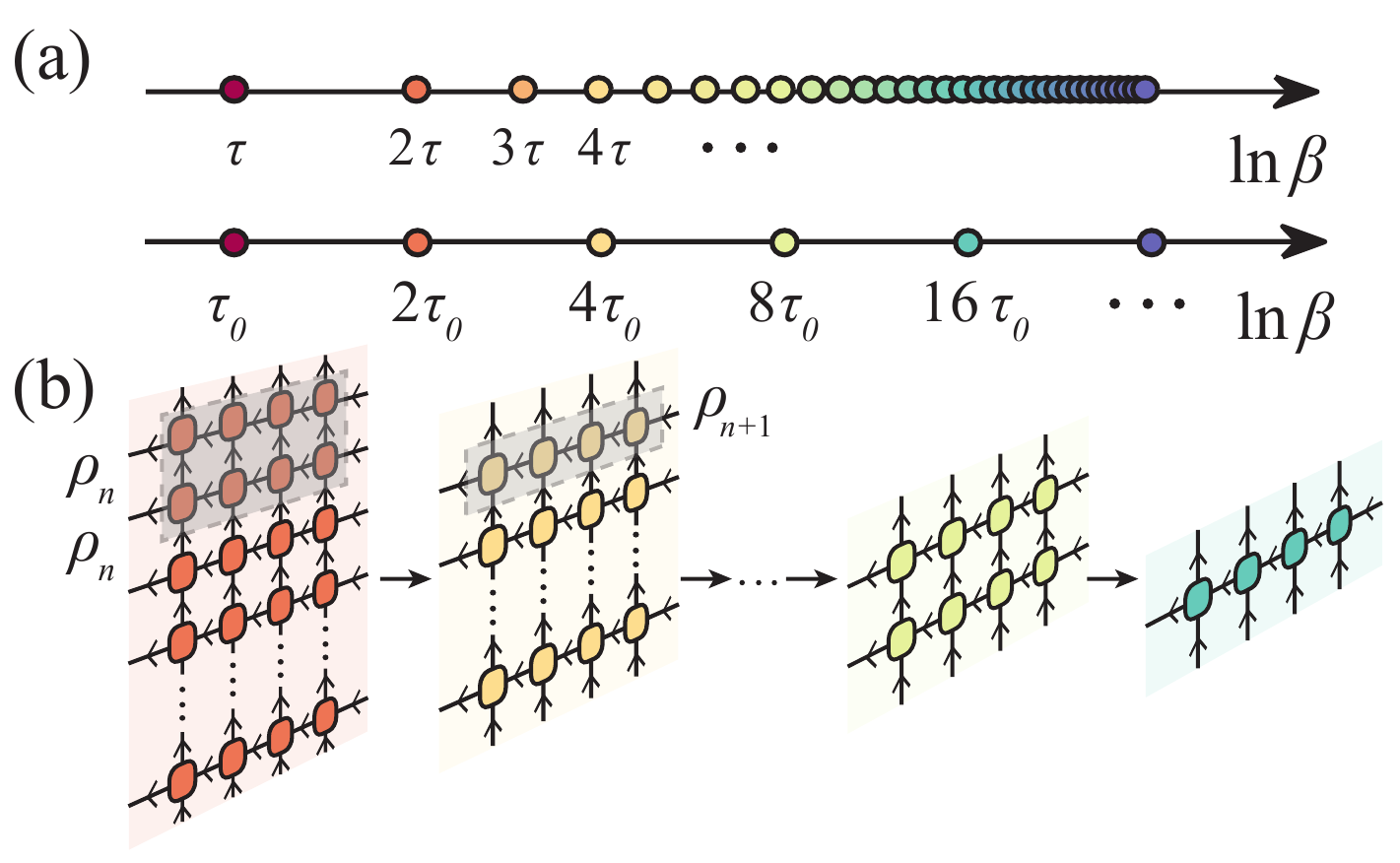}
\caption{(Color online)
 (a) Exponential vs. linear thermal evolution of the
 unnormalized density operator, $\rho \equiv e^{-\beta H}$.
 (b) The procedure of MPO doubling in XTRG, i.e.
 $\rho_{n+1} = \rho_n \ast \rho_n$ corresponding to
 $\beta_{n+1} =2\beta_n = 2^n \tau_{0}$.
}
\label{Fig:XTRG}
\end{figure}

We start from the MPO-based XTRG approach, 
which was proposed in Ref.~\cite{Chen2018}. 
There the (unnormalized) thermal mixed
state $\rho({\beta}) \equiv e^{-\beta H}$ of the system is
represented as 1D MPO, as depicted in Fig.~\ref{Fig:XTRG}(b).
In the purification framework, $\rho({\beta})$ represents a
thermo-double field, where the trace with its conjugate results
in the partition function at $2\beta$, i.e.,
\begin{equation}
Z(2 \beta) = \rm{Tr} [\rho(\beta) \cdot \rho^{\dagger}(\beta)].
\label{Eq:Z}
\end{equation}
Similarly, one can compute thermodynamic quantities. While
equivalent to the concept of purification, we emphasize,
however, that here we always describe the density matrix as an
MPO and thus as an operator. The added benefit of using
$\rho\rho^{\dagger}$ as in \Eq{Eq:Z} is that it always yields a
positive and, thus, a physical thermal density operator at
inverse temperature $2\beta$.

The evolution of $\rho(\beta)$ always starts from infinite
temperature in thermal RG approaches, where the density operator
is just a trivial identity, i.e. $\rho(0) = \mathbb{I}$.
Conventional linear RG approaches evolve the density matrix
linearly in $\beta$, i.e.,
\begin{equation}
\rho(n\tau) = \underset{n}{\underbrace{\rho(\tau) \cdot \rho(\tau)
\cdot ... \cdot \rho(\tau)}} \cdot I \,,
\end{equation}
where $\tau$ is a small imaginary time (inverse temperature)
step size that applies to all $\beta_{n+1} = \beta_n+\tau$.
We refer to such a scheme, schematically depicted
in the upper row of Fig.~\ref{Fig:XTRG}(a),
as a linear thermal RG, see Ref.~\cite{Li.w+:2011:LTRG}.

A recent insight from the logarithmic
entanglement scaling in
conformal field theory proves that the
block-entanglement growth in
a thermal MPO is upper bounded by
$S_E \leq a \ln \beta + b$ with some constants $a$ and $b$
\cite{Barthel.t:2017:FiniteT,Dubail17}. This shows that the
entanglement entropy $S_E$ of the MPO actually changes
significantly only when $\beta$ changes by a factor. In light
of this, the linear evolution is realized to be a very slow
cooling procedure, while a more efficient way is to follow the
logarithmic temperature scale. A particularly simple and
convenient choice is $\beta \to 2\beta$, i.e.,
\begin{equation}
   \rho_{n+1} = \rho_{n} \ast \rho_{n}, 
\label{Eq:XTRG}
\end{equation}
with $\rho_n \equiv \rho(2^n \tau_{0})$, and $\tau_0$ an
arbitrarily small {\it initial} starting point.
The asterisk here is a reminder to emphasize the underlying MPO
product structure. This exponential
procedure is illustrated in lower row of \Fig{Fig:XTRG}(a),
as well as in Fig.~\ref{Fig:XTRG}(b), which reveal manifestly
the efficiency of XTRG, i.e., the system reaches the lowest
temperature exponentially fast. Similarly, by its very construction,
XTRG can also start from an arbitrarily small $\tau_0$, which
thus allows one to resort to a very simple but accurate
initialization of $\rho_0$, e.g. using series expansion,
\begin{equation}
\rho_0 \equiv \rho(\tau_0)
= \sum_{n=0}^{n_c} \frac{(-\tau_0)^n}{n!} H^n,
\label{Eq:SETTN}
\end{equation}
even using a cutoff as small as $n_c=1$ where the MPO for
$\rho_0$ is given -- up to a very minor adaptation -- by the MPO
of $H$ itself with the {\it same} bond dimension. For
comparison, $n_c=2$ only includes one further doubling of the
MPO of $H$ with itself, an elementary MPO procedure in XTRG
in any case. Overall, the techniques required to perform the
expansion in Eq.~(\ref{Eq:SETTN}) have been developed in
Ref.~\cite{Chen.b+:2017:SETTN}, dubbed as series-expansion
thermal tensor network (SETTN) method.

Naively, one might expect that the numerical cost for the
step in \Eq{Eq:XTRG} scales like 
$\mathcal{O}(D^6)$ with $D$ the bond dimension of MPO,
and thus represents a (prohibitively) expensive calculation.
The numerical cost, however, can be strongly
reduced to 
$\mathcal{O}(D^4)$ 
by resorting to 
a variational
procedure (see \cite{Chen.b+:2017:SETTN} for details),
thus allowing 
larger $D$ in the calculation.
In addition, thanks to the versatile QSpace
framework \cite{Weichselbaum.a:2012:QSpace}, we have fully
implemented non-Abelian symmetries in our XTRG simulations
which effectively reduces the bond
dimension by switching from $D$ states to $\Dstar$ multiplets.
Conversely, for the case of SU(2) in the present study, 
we can therefore keep up to
$D \lesssim 4 \times \Dstar$ individual U(1) states.

Finally, we would like to emphasize that XTRG is not only
superior to conventional linear RG in efficiency, but also
in terms of accuracy. Since the lowest temperature
can be reached in an extremely speedy fashion,
this results in a significantly smaller number of numerical iterations.
Hence also the truncation error
accumulation is greatly reduced. A detailed comparison
of accuracies and efficiencies between XTRG and LTRG, as well as
SETTN, also can be found in Ref.~\cite{Chen2018}.


\subsection{Tensor Product Operator Approach}
\label{Sup:TPO}

As an alternative approach in our simulations,
we also utilize the tensor product
operator (TPO) approach, 
whose results generally are in agreement with XTRG.
Here we provide details of our TPO algorithm
together with more benchmark calculations.
For practical and historical reasons, we still use
a numerical code that is based on
a linearized scheme in the imaginary time evolution based
on Trotter decomposition in our TPO algorithm,
and also does not yet exploit any symmetries. Its
concepts, nevertheless, can equally well be generalized
in the spirit of doubling of the density matrix as in XTRG.

\begin{figure}[tbp]
\includegraphics[angle=0,width=0.5\linewidth]{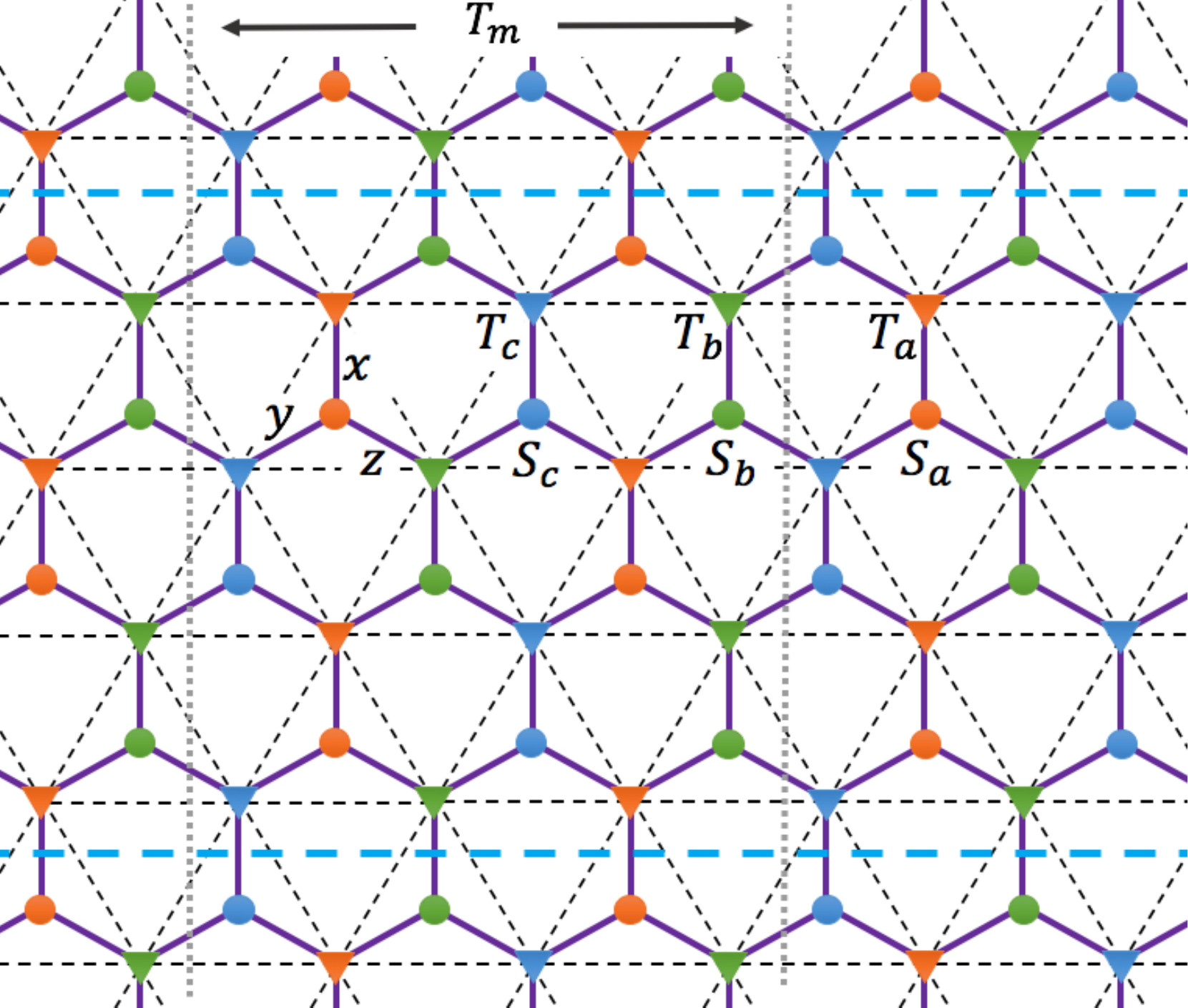}
\caption{(Color online)
   The triangular lattice (dashed lines) represented in
   the TPO ansatz (thick lines with symbols for tensors).
   Since we use Trotter decomposition within our TPO
   simulations, we need to differentiate sets of three tensors.
   There are three kinds of rank-5
   tensors, i.e. tensors with
   five external indices, namely $T_a$, $T_b$ and $T_c$
   (orange, green, and blue triangles, respectively)
   located on the sites of the TLH,
   where the two physical indices for bra and ket, i.e.
   $\sigma$ and $\sigma'$, are
   omitted for simplicity. There are also three kinds of
   rank-3 intermediate tensors $S_a$, $S_b$ and $S_c$
   (orange, green, and blue circles, respectively),
   without physical indices, residing in the center of the face of
   the up-triangles in the TLH.
   The geometrical bonds can be grouped by their orientation,
   labelled $x$, $y$ and $z$.
   The two horizontal dashed blue lines
   denote the XC boundary condition,
   where the system is wrapped up vertically
   (having $W=4$ in the plot above).
   The two vertical lines indicate the
   transfer matrix $T_m$ along the cylinder
   where, by definition of a transfer matrix, all
   `internal' indices, including the
   physical indices as well as intermediate
   geometric indices, should be traced,
   with the only open indices the thick lines
   crossing the vertical dashed lines.
}
\label{Fig:TPO}
\end{figure}

\sparagraph{TPO Representation of the Density Matrix}

From a numerical efficiency point of view,
it is favorable to describe the thermal states of TLH
via an effective hexagonal lattice TN,
as shown in \Fig{Fig:TPO}.
Furthermore, given that the TLH carries a $120^\circ$
magnetic order at zero temperature, this naturally
divides the sites on the TLH into three sublattices,
which is also required for a Trotter decomposition
in any case.
Therefore we introduce three types of $T$ tensors, 
$T_a$, $T_b$, and $T_c$ on the $A$-, $B$-, and $C$-sites
(orange, green and blue solid triangles in \Fig{Fig:TPO},
respectively). All of them are rank-5 tensors
with 2 physical indices and 3 geometric ones.
The $T$ tensors are interconnected to
form the hexagonal tensor network (TN) in \Fig{Fig:TPO}
along the $x$-, $y$-, $z$-bonds via
three types of rank-3 $S$ tensors, 
$S_a$, $S_b$ and $S_c$
(orange, green and blue solid circles, respectively)
with no physical indices of their own.
The $S$ tensors reside on three distinct up-triangles,
dubbed as $a$-, $b$-, and $c$-triangles,
corresponding to the three sublattices $A$, $B$, and $C$,
respectively.
They contain disconnected subsets of nearest-neighbor
terms on the TLH which can be readily utilized for a Trotter
decomposition [cf. \Eq{Eq:DenMat}].
In \Fig{Fig:TPO} we also indicate
how to wrap up the TPO on an XC, i.e.,
by imposing periodic boundary condition (BC) along the vertical direction.

Throughout, we assume that the (starting)
configuration of the $S$ 
tensors is in the canonical form,
\begin{equation}
\label{Eq:SProperty}
\sum_{i, j} (S_\alpha)^*_{ijk}(S_\alpha)_{ijk^{\prime}} = (\lambda^{\alpha}_{k})^2 \delta_{kk^{\prime}}.
\end{equation}
where the $\alpha\in a,b,c$ 
labels different $S$ tensors,
and the bond indices $(i,j,k)$ denotes
cyclic permutations of $(x,y,z)$.
This form can be achieved by higher-order
singular-value decomposition
(HOSVD), with a residual unitary out of the
$S$ tensors absorbed
into the $T$ tensors, and the singular values
$\lambda^{\alpha}_k \ge0$
reabsorbed into the $S$ tensor.
The contraction in \Eq{Eq:SProperty}
thus regenerates a diagonal matrix with
entries $(\lambda^{\alpha}_k)^2$.
For a more detailed discussion of HOSVD,
we refer the reader to Refs.~\cite{Lathauwer2000,Xie2014}

\begin{figure}[!tbp]
\includegraphics[angle=0,width=0.85\linewidth]{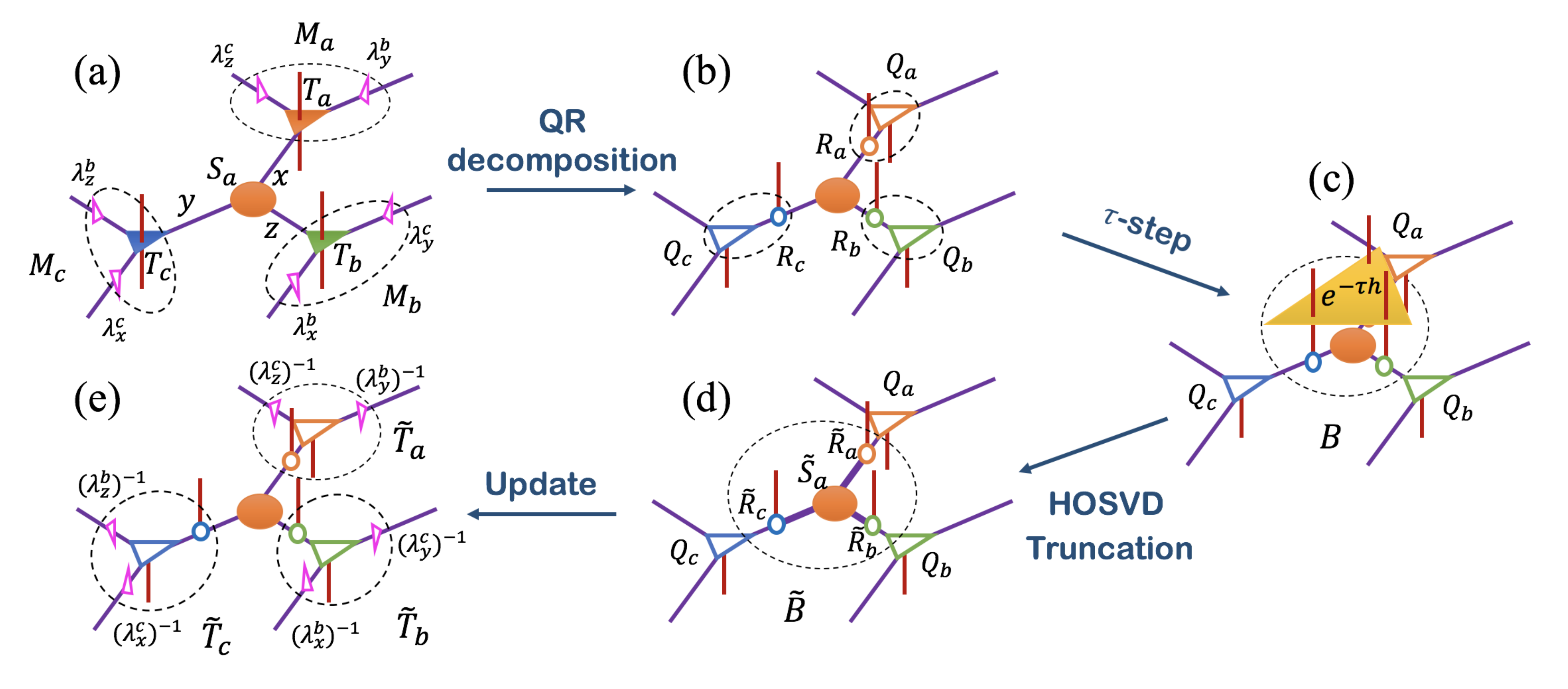}
\caption{
    (Color online) Procedure of a single projection step:
(a) Absorptions of
    $\lambda$ matrices, where the suggestive arrow with
    the $\lambda$'s indicates the (approximate)
    orthonormalization
    of the entire Bethe TNS.
(b) QR decomposition, while also splitting off the upper
    physical indices.
(c) Applying the triangular imaginary time step
    $e^{-\tau h}$ to the base tensor.
(d) HOSVD of the base tensor.
(e) truncate and update $S_a$ as well as three $T$ tensors.
}
\label{Fig:Projection}
\end{figure}

\sparagraph{Imaginary-time Evolution and Simple Update Scheme of TPO}

We utilize the Bethe-lattice approximation, also referred
to as simple update \cite{Jiang2008,Li2012},
in the imaginary-time evolution to optimize
the TPO density matrix of the system from high to low
temperatures. To be concrete, via a Trotter-Suzuki
decomposition, the density matrix can be expressed as
\begin{eqnarray}
    \rho = e^{-\beta H} \approx
    \bigl(e^{-\tau H^A}e^{-\tau H^B}e^{-\tau H^C}\bigr)^N \equiv
    \bigl(P^A P^B P^C\bigr)^N \quad \text{with} \quad 
    P^\Lambda \equiv \prod_{i\in \Lambda} e^{-\tau h_{i}}, 
\label{Eq:DenMat}
\end{eqnarray}
where $H^\Lambda \equiv \sum_{i\in\Lambda} h_i$,
with 
$h_i$ 
the `triangular plaquette' Hamiltonian
on the up-triangle $i$ in sublattice $\Lambda\in A,B,C$,
and $\tau$ the Trotter step (chosen as $0.01\sim0.02$ in practice).
By construction, $[h_{i}, h_{i'}]=0$ 
for $i\ne i'$ within the same sublattice
$i,i' \in\Lambda$. All of the $h_i$ 
have identical form
for the isotropic Heisenberg model considered here.
We initialize with an infinite-temperature density matrix
(direct product of identities), and apply the triangular operators
$P^{\Lambda}$ 
sequentially for $\Lambda \in A,B,C$.
This is repeated iteratively to cool down the TPO density matrix.

To be specific, we now describe in detail
the application of $P^A$,
where the tensor $S_a$ is surrounded by the
tensors $T_a$, $T_b$ and $T_c$
(see \Fig{Fig:TPO} and also \Fig{Fig:Projection}).
\begin{enumerate}
\renewcommand{\labelenumi}{(\roman{enumi})}
\renewcommand{\theenumi}{\roman{enumi}}

\item We firstly (re)generate the diagonal
$\lambda^\alpha$ matrices as in \Eq{Eq:SProperty}
out of the $S_\alpha$ tensors surrounding the $T$ tensors, \\
i.e. $\alpha \in \{b,c\}$.

\item \label{item:2}
Next we absorb the $\lambda$'s
into the $T$'s to construct the tensors $M$
[\Fig{Fig:Projection}(a)]
\begin{subequations}
\label{Eq:AbsorbLam}
\begin{eqnarray}
	(M_a)^{\sigma_a'\sigma_a}_{xyz} &=&(T_a)^{\sigma_a'\sigma_a}_{xyz} \lambda_{y}^b \lambda_{z}^c, \\
	(M_b)^{\sigma_b'\sigma_b}_{xyz} &=& (T_b)^{\sigma_b'\sigma_b}_{xyz} \lambda_{x}^b \lambda_{y}^c, \\
	(M_c)^{\sigma_c'\sigma_c}_{xyz} &=& (T_c)^{\sigma_c'\sigma_c}_{xyz} \lambda_{z}^b \lambda_{x}^c
\end{eqnarray}
\end{subequations}
(note that we explicitly indicate summation, i.e. there is no
implicit summation here over double indices since the $\lambda$
matrices are diagonal).

\item Perform QR decompositions of $M_a$, $M_b$ and $M_c$
[\Fig{Fig:Projection}(b)] \label{item:3}
\begin{subequations}
\label{Eq:QRc}
\begin{eqnarray}
	(M_a)^{\sigma_a'\sigma_a}_{x^{\prime}yz} &=& \sum_{x} (Q_a)^{\sigma_a'}_{xyz} (R_a)^{\sigma_a}_{xx^{\prime}}, \label{Eq:QRa}\\
	(M_b)^{\sigma_b'\sigma_b}_{xyz^{\prime}} &=& \sum_{z} (Q_b)^{\sigma_b'}_{xyz} (R_b)^{\sigma_b}_{zz^{\prime}}, \label{Eq:QRb} \\
	(M_c)^{\sigma_c'\sigma_c}_{xy^{\prime}z} &=& \sum_{y} (Q_c)^{\sigma_c'}_{xyz} (R_c)^{\sigma_c}_{yy^{\prime}}.
\end{eqnarray}
\end{subequations}
Here we have also
split off the upper physical indices into the tensors
$R_a$, $R_b$ and $R_c$
which significantly reduces the computational
cost in the projection-truncation procedure in the next step.

\item Construct the base tensor $B$ by contracting $R_a$, $R_b$,
$R_c$ with $S_a$ tensor
\begin{subequations}
\begin{equation}
\label{Eq:BaseTens}
	B^{\sigma_a\sigma_b\sigma_c}_{xyz} = \sum_{x^{\prime},y^{\prime},z^{\prime}} (R_a)^{\sigma_a}_{xx^{\prime}} (R_b)^{\sigma_b}_{zz^{\prime}} (R_c)^{\sigma_c}_{yy^{\prime}} (S_a)_{x^{\prime}y^{\prime}z^{\prime}},
\end{equation}
and apply the 3-site imaginary-time step $P=e^{-\tau h}$
onto the base tensor [\Fig{Fig:Projection}(c)]
\begin{equation}
\label{Eq:Projection}
	\tilde{B}^{\sigma_a\sigma_b\sigma_c}_{xyz} = \sum_{\sigma^{\prime}_a,\sigma^{\prime}_b,\sigma^{\prime}_c}
	P^{\sigma_a\sigma_b\sigma_c}_{\sigma^{\prime}_a\sigma^{\prime}_b\sigma^{\prime}_c} \cdot
	B^{\sigma^{\prime}_a\sigma^{\prime}_b\sigma^{\prime}_c}_{xyz}.
\end{equation}
\end{subequations}

\item Take HOSVD of the modified base tensor $\tilde{B}$
by performing independent SVD w.r.t. the index pairs
$(\sigma_a,x)$, $(\sigma_b,z)$, and $(\sigma_c,y)$,
providing the isometries $\tilde{R}_a$ to $\tilde{R}_c$,
respectively [\Fig{Fig:Projection}(d)]. Then by projecting the original tensor
$\tilde{B}$ with $\tilde{P}_\alpha \equiv
\tilde{R}_\alpha \tilde{R}_\alpha^\dagger$
on all three indices $\alpha \in \{a,b,c\}$, one obtains
the updated tensor $\tilde{S}$,
\begin{subequations}
\begin{equation}
\label{Eq:HOSVD-1}
	(\tilde{S}_a)_{xyz} =
    \sum_{\substack{ \sigma_a \sigma_b \sigma_c \\ x'y'z'} }
      (\tilde{R}_a^\ast)^{\sigma_a}_{x' x}
      (\tilde{R}_b^\ast)^{\sigma_b}_{z' z}
      (\tilde{R}_c^\ast)^{\sigma_c}_{y' y}
      \tilde{B}_{x'y'z'}^{\sigma_a \sigma_b \sigma_c},
\end{equation}
or equivalently,
\begin{equation}
\label{Eq:HOSVD}
	\tilde{B}^{\sigma_a\sigma_b\sigma_c}_{xyz} =
    \sum_{x^{\prime},y^{\prime},z^{\prime}}
      (\tilde{R}_a)^{\sigma_a}_{xx^{\prime}}
      (\tilde{R}_b)^{\sigma_b}_{zz^{\prime}}
      (\tilde{R}_c)^{\sigma_c}_{yy^{\prime}}
      (\tilde{S}_a)_{x^{\prime}y^{\prime}z^{\prime}}.
\end{equation}
\end{subequations}
In exact numerics, the bond dimensions of
$\tilde{S_a}$ would be generally
enlarged by the local Hilbert space dimension $d$, and
thus needs to be truncated.
This is achieved by discarding the smallest singular
values in $\lambda$, 
such that $\tilde{P}_\alpha$ becomes a true projector,
namely to the sector of dominant singular values.
The kept singular values are the ones that also occur
in \Eq{Eq:SProperty}. Without symmetry breaking,
the three directions $x,y,z$ are equivalent, and hence
the $S$ and $T$ tensors may be chosen symmetric under
cyclic permutation of these indices, throughout. This
simplifies the TPO step above in that only a single SVD
already suffices to obtain $\tilde{R} \equiv \tilde{R}_a =
\tilde{R}_b = \tilde{R}_c$. In practice, the results
were equivalent whether or not this lattice symmetry
was enforced.

\item The truncated $\tilde{R}$ tensors in \Eq{Eq:HOSVD}
can now be contracted (absorbed) into the $Q$ tensors
in (\ref{item:3}).
By also undoing step (\ref{item:2}) by applying inverted
$\lambda^b$ and $\lambda^c$ weights
(note that in the above steps,
only $\lambda^a_\xi$ with $\xi \in \{x,y,z\}$ was altered,
but the sets $\lambda^b_\xi$ and $\lambda^c_\xi$
remained the same), we obtain the updated
$\tilde{T}$ tensors [\Fig{Fig:Projection}(e)],
\begin{subequations}
\label{Eq:NewT}
\begin{eqnarray}
	(\tilde{T}_a)^{\sigma_a'\sigma_a}_{x^{\prime}yz}
  &=& \sum_{x} (Q_a)^{\sigma_a'}_{xyz} (\tilde{R}_a)^{\sigma_a}_{xx^{\prime}}
  / (\lambda_{y}^{b} \lambda_{z}^c), \\ 
	(\tilde{T}_b)^{\sigma_b'\sigma_b}_{xyz^{\prime}}
  &=& \sum_{z} (Q_b)^{\sigma_b'}_{xyz} (\tilde{R}_b)^{\sigma_b}_{zz^{\prime}}
  / (\lambda_{x}^{b} \lambda_{y}^c), \\ 
	(\tilde{T}_c)^{\sigma_c'\sigma_c}_{xy^{\prime}z}
  &=& \sum_{y} (Q_c)^{\sigma_c'}_{xyz} (\tilde{R}_c)^{\sigma_c}_{yy^{\prime}}
  / (\lambda_{z}^{b} \lambda_{x}^{c}). 
\end{eqnarray}
\end{subequations}

\end{enumerate}

\sparagraph{Evaluation of thermal quantities}

The $S$ and $T$ tensors from the simple update above are inserted
into the 2D-TN of \Fig{Fig:TPO}.
One then needs to contract this TN
efficiently in the thermodynamic limit to obtain 
the partition function, and thus physical
thermal properties such as free energy, energy, magnetization, etc.
This constitutes another
essential challenge of the algorithm.
For finite-size cylinders with a small width $W$, e.g., XC4,
we perform exact contractions;
while for an infinite-size system,
we use conventional boundary
matrix product state (MPS) technique adopted in infinite
projected entanglement pair state (iPEPS) algorithms
\cite{Jordan2008}.

For both the XC4 and infinite-size systems, 
the dominating eigenvector
as well as eigenvalue of the horizontal 
transfer matrix $T_m$
(i.e. with a cut in Y direction, see Fig.~\ref{Fig:TPO})
can be obtained exactly (for the XC4) or approximately 
(for infinite systems)
by iteratively contracting
a trial initial vector with $T_m$ until convergence.
For this, double-layer TPO, i.e. using
$\rho(\beta/2)^\dagger \rho(\beta/2) \rightarrow \rho(\beta)$
to enforce positivity, is orders of magnitude more
expensive as compared
to the relatively cheap single-layer formalism of TPO.
For the XC4 system, 
the single-layer TPO computations are affordable
at a cost of $O(D^6)$,
where $D$ up to 60 is the bond dimension of TPO.
For the subsequent embedding into an infinite 2D TNS,
the TPO cost scales as $O(\chi^3 D^3)$, 
with $\chi$ the bond dimension of the boundary MPS.
In practice, we choose $D$ up to 40 and $\chi \sim 4D$, 
which is affordable, yet also
ensures data convergence over the parameter $\chi$.

\begin{figure}[htbp]
\includegraphics[angle=0,width=0.8\linewidth]{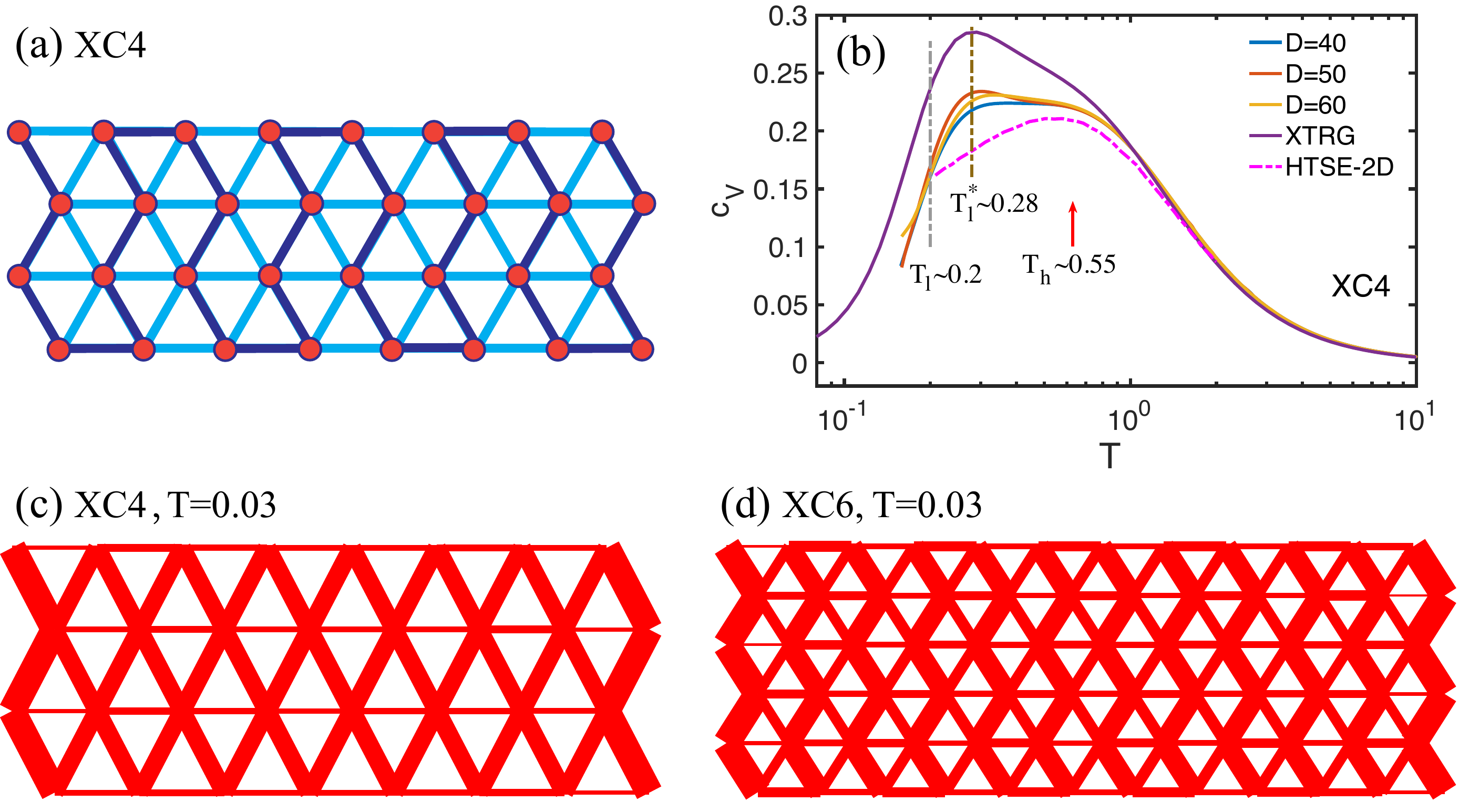}
\caption{(Color online)
   (a) XC geometry where the dark blue
   lines indicate the MPO path in
   the XTRG simulations.
   (b) Comparison of specific heat $c_V$ between
   TPO and XTRG on XC4 and, for reference,
   also HTSE-2D, i.e.,
   not constrained to the XC4 geometry.
   The high temperature scale,
   $\Th\sim0.55$, indicates the high temperature round peak,
   while the low temperature scale, $\Tl^* \sim 0.28$,
   is slightly higher than $\Tl \sim 0.20$
   for wider systems, e.g., YC6 in
   the main text as well as XC6 below.
(c,d) illustrate the bond energy textures on XC4 (no vertical stripes)
and XC6 (vertical stripes) geometries,
consistent with $T=0$ DMRG simulations \cite{Wb11incomm}.
 The explicit symmetry breaking in the vertical direction
 by forming stripes, here in (d), at finite temperature is necessarily
 linked to the choice of the open boundary edges.
}
\label{Fig:XC4Cv}
\end{figure}

\section{Triangular Lattice Antiferromagnet}
\label{Sup:XC}
\subsection{Benchmark results on XC4 geometry}
\begin{figure}[tbp!]
\centering
\includegraphics[angle=0,width=0.6\linewidth]{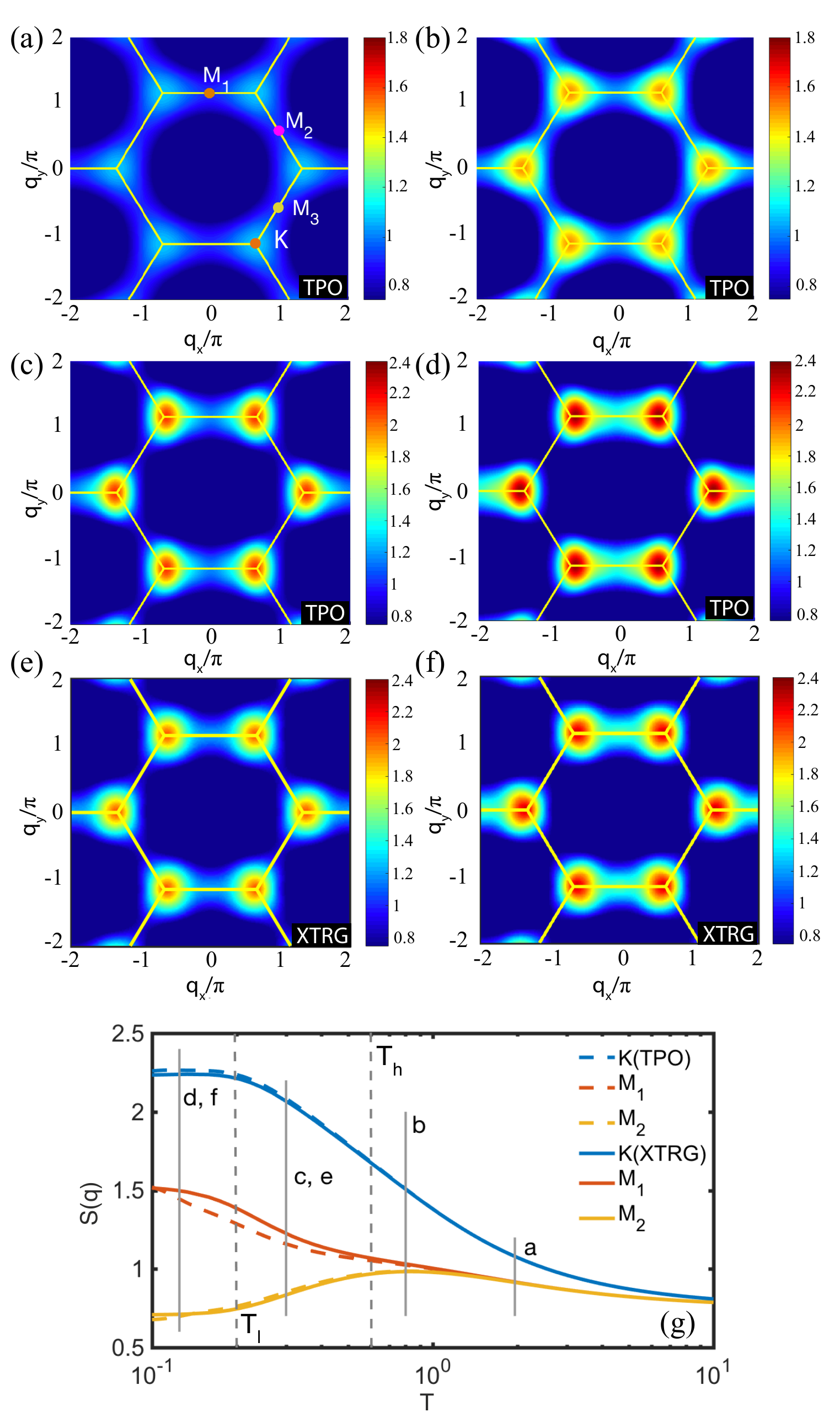}
\caption{(Color online) Static structure factor $\Sq$ of XC4,
     i.e., with $q_x$ pointing along the cylinder,
     computed by TPO (a-d) and XTRG (e,f)
     at various temperatures, $T\simeq 2,0.8,0.3,0.125$,
     as also indicated by the vertical solid gray lines in (g).
     While only having $W=4$ discrete $q_y$ momenta due to
     periodic BC, we interpolate by adding
     further intermediate points $q_y$
     in the Fourier transform in $S(q)$ to obtain 
     for overall smoother results.
     The vertical dashed lines in (g) indicate \Tl and \Th.
      Overall, the TPO data in (c,d) shows good agreement with
      the XTRG data in (e,f) at the same respective temperatures.
    (g) $\Sq$ at $q=K$ and two types of $M$ points [cf. panel (a)]
    vs. $T$, again with good quantitative agreement between
    TPO and XTRG.
    \vspace{-0.2in} 
}
\label{Fig:XC4SF}
\end{figure}

In contrast to YC geometries
which correspond to
a cylinder with straight ends, the XC geometry results
in a cylinder with zigzag ends \cite{Wb11incomm}.
Also note that the YC systems have a unique shortest
distance path along the nearest-neighbor bonds of the TLH
around the circumference when starting
from an arbitrary but fixed lattice site. This can favor
1D RVB stripe structures around the circumference, in contrast
to XC systems. Here therefore 
we apply both methods above to an XC4 geometry,
i.e., of width $W=4$,
as shown in \Fig{Fig:XC4Cv}(a).
For XTRG, we use our default 
aspect ratio $L/W=2$, and map the 2D lattice into a 1D structure
along the snake line also shown in Fig.~\ref{Fig:XC4Cv}(a).
For the TPO method, 
we optimise the tensors $T$ and $S$
directly in the thermodynamic limit, and connect the local
tensors on an infinitely long XC4 lattice.

In Fig.~\ref{Fig:XC4Cv}(b), the specific heat, $c_V$, of
XC4 obtained by XTRG by retaining $\Dstar=400$
multiplets, and by TPO with different bond dimensions
$D$ up to 60, is presented in comparison with HTSE.
The two temperature scales are apparent in both cases,
although the low temperature peak appears at $\Tl^*\sim0.28$
somewhat above the value $0.20$ obtained
on wider YC geometries.
This is due to strong finite-site effects on XC4,
which will be further analyzed
shortly by calculating the static structure factor $\Sq$.
Another distinct feature 
is that the lower-temperature peaks
in TPO curves are significantly 
lower than for the XTRG results.
This may also hint
a smaller finite-size effect in the TPO approach
(due to its simple update scheme).

In \Fig{Fig:XC4SF}(a-f),
we visualize the static structure factors $\Sq$ for XC4 in the first
BZ at various temperatures, which are marked
by grey solid lines in \Fig{Fig:XC4SF}(g)
together with the corresponding
panel reference [(a-d) from TPO,
and (e,f) from XTRG calculations].
When lowering the temperature,
the triangular lattice symmetry
is broken around \Th due to the finite system size, as manifested
by the slightly different behavior 
of the data for the otherwise equivalent
points $M_1$ and $M_2$ [indicated by markers in \Fig{Fig:XC4SF}(a)].
For temperatures below \Th,
one can observe that
$S(M_1)$, which points perpendicular to the direction
of the cylinder, turns brighter whereas $S(M_2)$
and $S(M_3)$ start loosing weight.
This indicates a tendency for
enhanced AF, i.e. N\'eel like correlations
around the circumference of the cylinder.
Note that for XC4 one has
equivalent, i.e. non-symmetry broken
shortest zig-zag
paths around the circumference of the cylinder [vertical
direction in \Fig{Fig:XC4Cv}(c)].

In \Fig{Fig:XC4SF}(g) we also directly compare $\Sq$ vs. $T$
at $M_{1,2}$ and $K$ from both methods, XTRG as well as TPO.
At the $K$ and $M_2$ points, the data from the two methods
coincide, while for $M_1$, the XTRG data
reaches the low-$T$ limit faster. This may be attributed
to a stronger (or cleaner) finite-size effect in XTRG,
as compared to the TPO data which, despite being evaluated
on a cylinder, originates from an infinite-system simple
update. A similar conclusion was already 
drawn from the data in \Fig{Fig:XC4Cv}(b) regarding 
the higher $c_V$ peak for XTRG at $\Tl^*$.

\begin{figure}[tbp!]
\includegraphics[angle=0,width=0.85\linewidth]{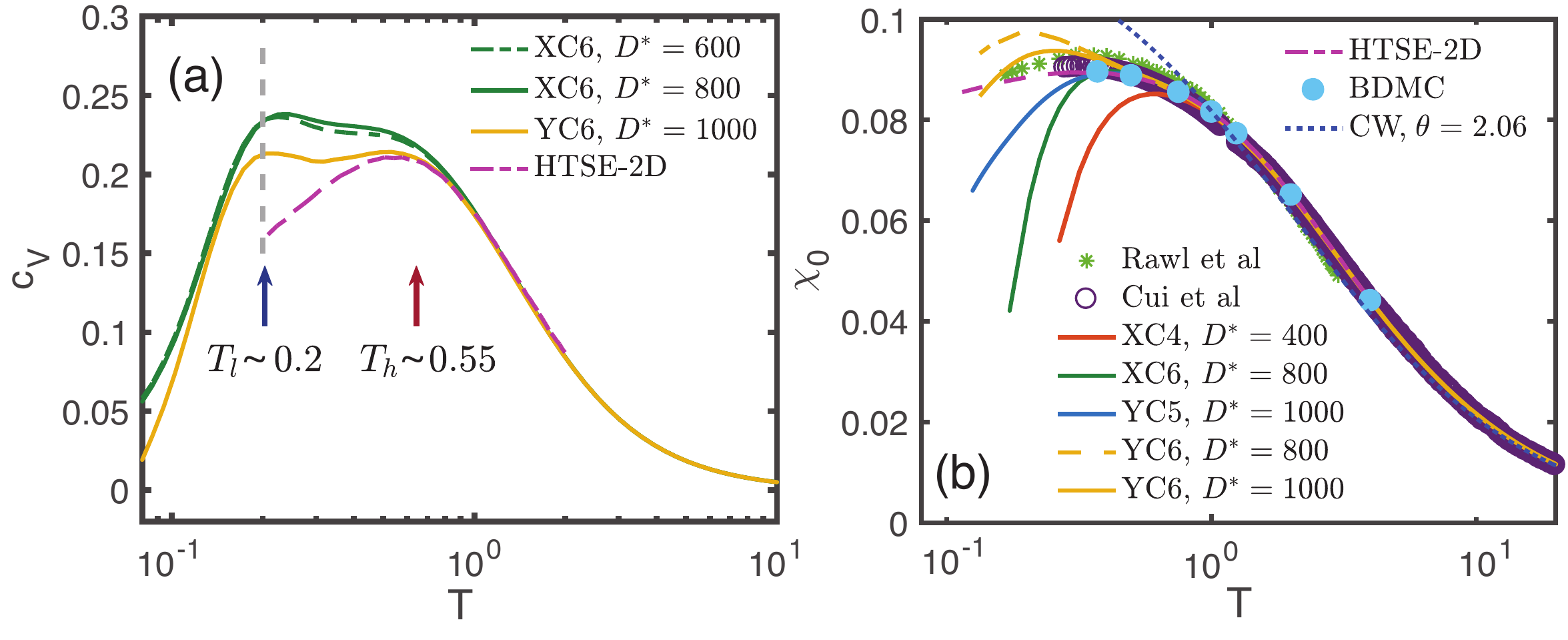}
\caption{(Color online)
(a) Specific heat $c_V$ on XC6 and YC6 lattices.
(b) Static magnetic susceptibility $\chi_0$ vs. $T$.
    We also added HTSE-2D data and other data from
    the literature for reference, as well as
    a simple Curie Weiss (CW) estimate (same as in  
    Fig.~2 in the main paper).
    For a discussion of numerical cost, i.e.
    the growth of the block entanglement $S_E$
    with decreasing $T$,
    see, e.g., Fig.~3(f) in the main text.
}\label{Fig:XC6}
\end{figure}

\begin{figure}[tbp!]
\includegraphics[angle=0,width=1\linewidth]{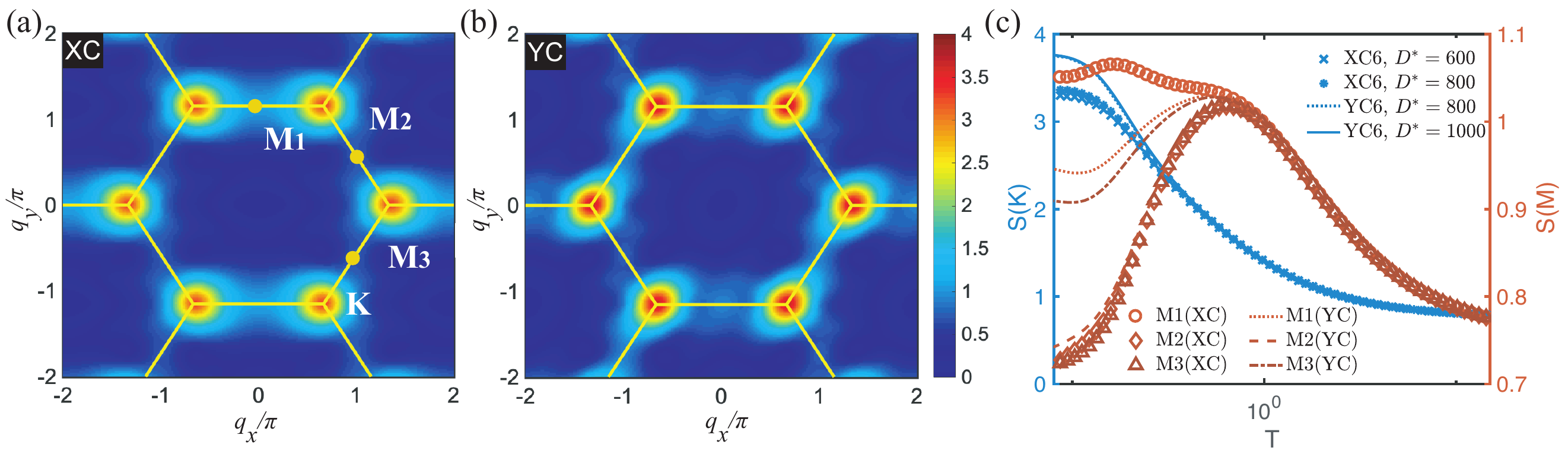}
\caption{(Color online)
  Comparison of structure factor between 
  (a) XC$6{\times}12$ and
  (b) YC$6{\times}12$ geometries at $T = 0.103$.
  The direction along the cylinder corresponds to $q_x$
  and $q_y$, respectively.
  XC6 shows a higher intensity around $M_1$
  which indicates a tendency to form stronger bonds
  around the circumference of the cylinder, 
  i.e., stripe order.
  Consequently, XC6 has a lower
  $S(K)$ as compared to YC6, shown more explicitly in (c).
}
\label{Fig:XC6SF}
\end{figure}

\subsection{Specific heat, susceptibility, and structure factor
for cylinders up to $W=6$}

We present our XTRG results on XC6 and YC6
systems for the specific heat and static
susceptibility in \Fig{Fig:XC6}, and for the static
structure factor in \Fig{Fig:XC6SF}.
Overall, we expect clearly reduced
finite-size effects as compared to the width $W=4$ systems.

The specific heat, $c_V$, on XC6, shown in
\Fig{Fig:XC6}(a), agrees well with both YC6 and HTSE in
the high temperature regime, $T\gtrsim \Th$.
The observed lower energy scale
in this data is stable around $\Tl\sim0.2$
also for this wider system.
In Fig. \ref{Fig:XC6}(b), we compare
our XTRG data for the magnetic
susceptibility, $\chi_0$ vs. $T$, on XC6 with
other results including two experimental measurements,
XC4, YC5 and 6, HTSE, etc.
A good agreement between XC6 data and experimental
results can be observed, although YC6 produces $\chi_0$ mostly
close to experiments and constitutes the overall most suitable 
geometry.

From Figs. \ref{Fig:XC6}(a,b)
we can conclude that although XC6
has many features in common with YC6,
it suffers larger
finite-size effects and is less favorable in approximating 
the thermodynamic limit. This can be directly
seen in the bond energy texture on XC6 in \Fig{Fig:XC4Cv}(d).
Although being measured at a finite temperature,
it is consistent with previous DMRG studies
of the ground state properties \cite{Wb11incomm}, where
spontaneous zig-zag stripe formation around the
circumference of the cylinder is found 
for XC systems of width $2n+4$, with $n$ an integer.

The structure factor, $\Sq$, is analyzed in \Fig{Fig:XC6SF},
where XC6 [\Fig{Fig:XC6SF}(a)] is compared to
YC6 [\Fig{Fig:XC6SF}(b)].
By reaching temperatures as low as
$T \sim 0.1 <\Tl$,
we expect signatures of incipient $120^\circ$ order
to be present in both systems.
For YC6 cases, there exist two $M$ points ($M_1$ and
$M_3$) which have strong intensity,
while the remaining one, $S(M_2)$, is weak
at low temperatures, although all three $M$ points have
anomalous enhancement at $T$ around $\Th$, 
due to RLE activation. In XC6, there is one
strong $S(M_1)$ as compared to weaker $M_2$ and $M_3$ points, 
indicating a strong tendency for stripe
order [see also Fig.~\ref{Fig:XC4Cv}(d)],
which is largely absent in YC6 case.
Consequently, the strength of the $K$-point correlations in XC6 is
impaired as well compared to the YC6 case [see Fig.~\ref{Fig:XC6SF}(c)].
The situation here is thus opposite to the width $W=4$ case,
where YC4 shows a stripe phase, whereas XC4 does not.

\begin{figure}
\includegraphics[angle=0,width=0.8\linewidth]{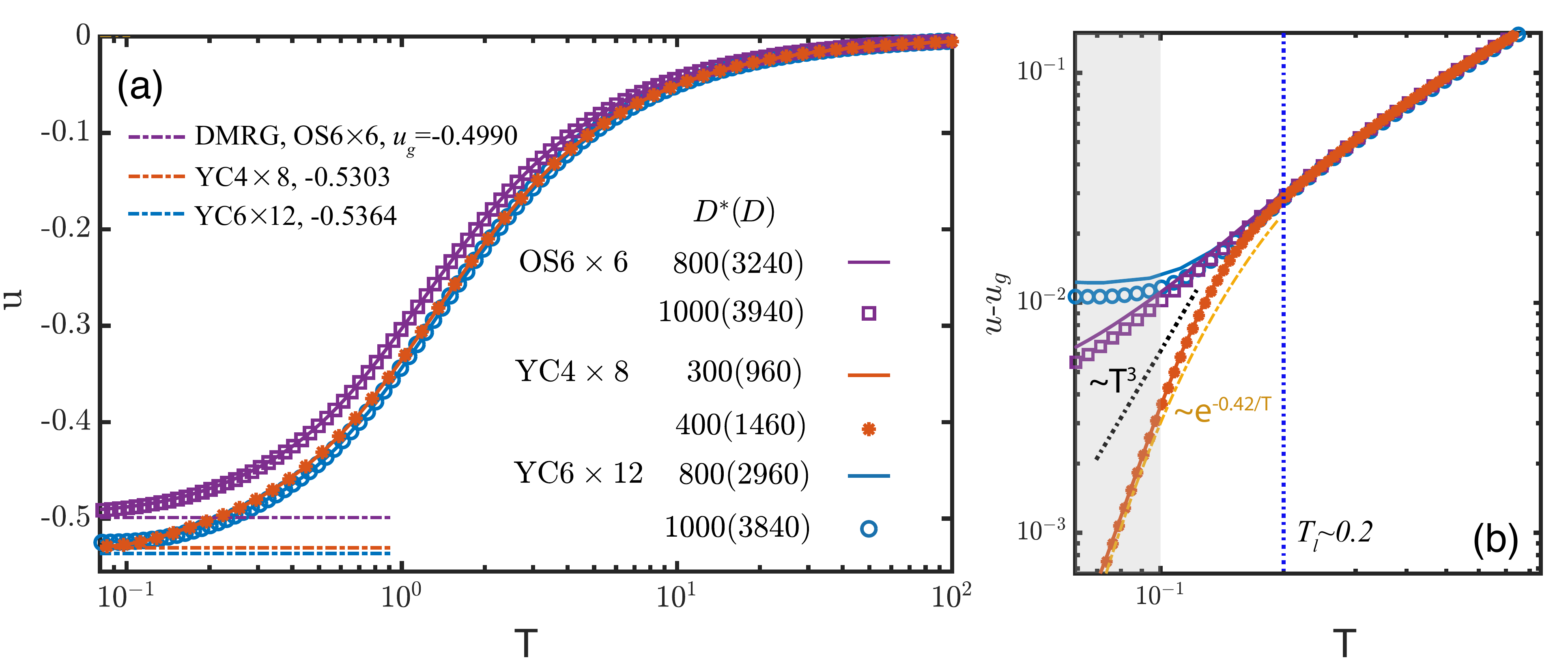}
\caption{
(a) Internal energy, $u$ vs. $T$, on
YC$6\times12$, OS$6\times6$, and YC$4\times8$ lattices.
The legend specifies the
number of multiplets $\Dstar$ followed by
the corresponding number of $U(1)$ states.
DMRG data $u_g$ (horizontal markers)
represent ground state results
where we kept up to $\Dstar=2000$ SU(2) multiplets.
(b) Same data as in (a)
but relative to $u_g$ on a log-log plot, from which
suggests algebraic convergence for OS6 and YC6.
The latter (YC6, blue), however,
is not yet fully converged below $T\lesssim 0.1$
(dashed regime) since it still changes with $\Dstar$.
In contrast, YC4 already shows
\textit{exponential} convergence
to the ground state energy 
for the lowest temperatures which may already
be attributed to the finite size energy gap,
as indicated by the exponential fitting.
The OS6 data (purple) exhibits a quasi-algebraic
behavior below $T_l$. The black dotted line
line is a guide to the eye and indicates the
expected $\propto T^{3}$ at low $T$ for
large enough systems [e.g., cf. \Fig{Fig:LSWT}].
}
\label{Fig:Es}
\end{figure}

\subsection{Internal energy}
In \Fig{Fig:Es}, we analyze the internal energy per site,
$u(T) \equiv \tfrac{1}{N}\langle H \rangle_T$ with
$c_V(T) \equiv \tfrac{\partial u}{\partial T}
= -\beta \tfrac{\partial u}{\partial \ln \beta}$ \cite{Chen2018},
versus temperature
and compare it to the ground state DMRG results on the same
geometry. We show data for various
geometries, including the
YC$6\times12$, OS$6\times6$, and small
YC$4\times8$ lattices.
It can be observed that the energy data are well
converged with bond
dimension $\Dstar=1000$ (for YC6 and OS6) and $\Dstar=400$ (YC4),
which correspond to nearly $D\sim 4\times\Dstar$ in terms of $U(1)$ states
(see legend).

At low temperature, the $u(T)$ curves already closely approach
the zero-temperature limits $u_g$ [horizontal markers
\Fig{Fig:Es}(a)] obtained by ground state density matrix
renormalization group (DMRG) calculations. For a strong
comparison, we replot the same XTRG data in \Fig{Fig:Es}(b), but
now relative to the DMRG ground state energy $u_g$ on a log-log
plot. The data for the intermediate to large temperature regime,
$T>\Tl$, scales similarly for all boundary conditions (with minor
offsets due boundaries). As $T$ decreases below $\Tl$, the YC4
data converges exponentially, in agreement with the
stripe-phase scenario revealed in Ref.~\cite{Chen2018}.
In contrast, the OS6 and YC6 data collapse onto each other down to
temperatures well below $\Tl$ (here the upturn in the YC6 system
(blue data) is attributed to finite-$D$ accuracy).
Note that from the OS6 data in \Fig{Fig:Es}(b),
one may estimate an
approximate power-law behavior.

\subsection{System size dependence of low-temperature scale $\Tl$}

\begin{figure}
\includegraphics[angle=0,width=0.55\linewidth]{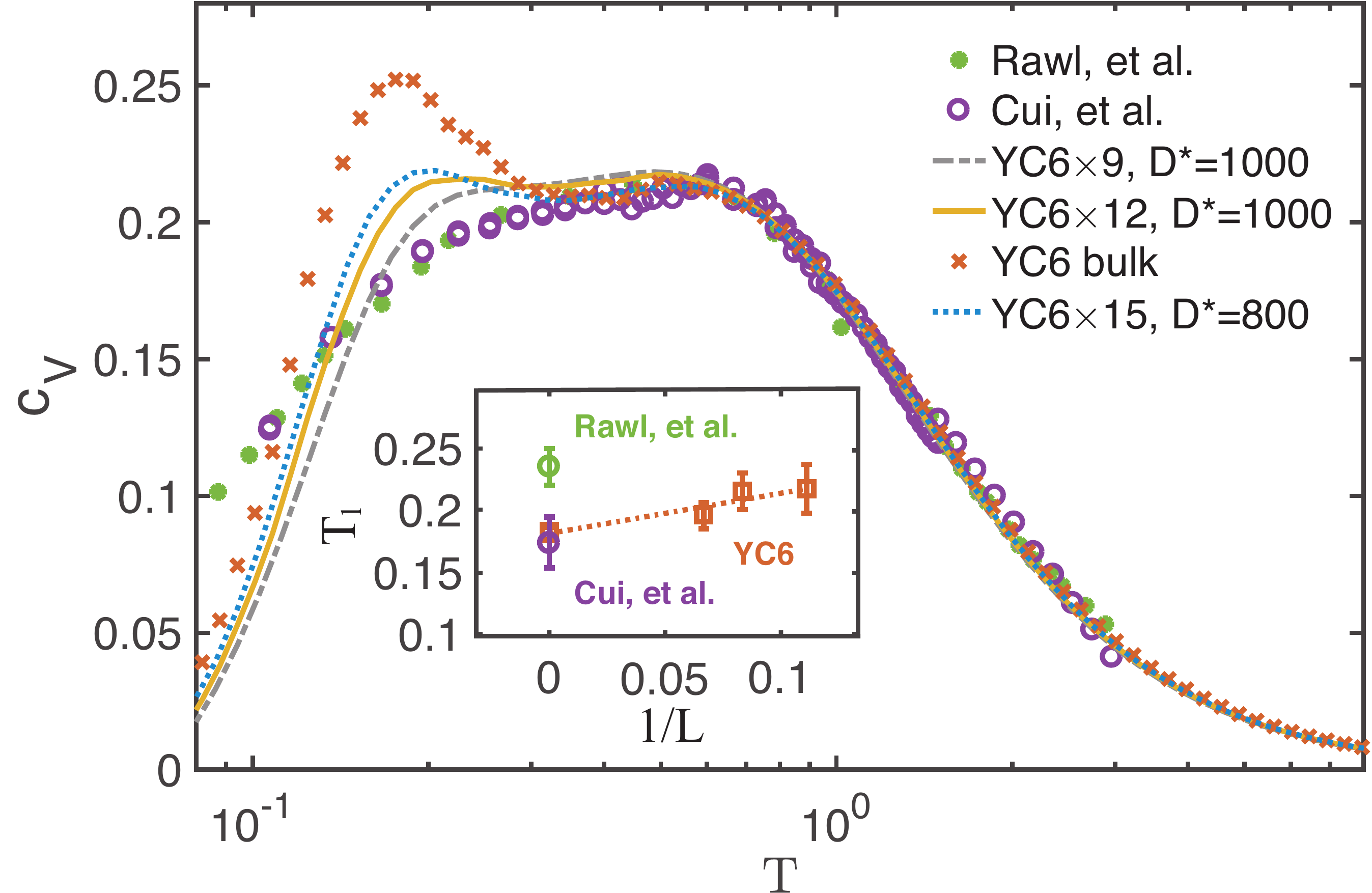}
\caption{
Specific heat on YC6 for three different lengths,
also compared to
experimental data. The `YC6 bulk' is obtained
by subtracting YC$6{\times}9$ from YC$6{\times}12$ data,
in order reduce the effects from the open boundary.
The inset shows the lower characteristic
temperature $\Tl$ vs. $1/$L with 
estimated error bars 
from the determination of $\Tl$ for 
the peak/shoulder structure.
The data point at $1/L=0$
is taken from `YC6 bulk' curve
which, indeed, approximately represents
an extrapolation in $1/L\to0$, as suggested by
the dotted line.
In particular, to determine $\Tl$ in experimental curves,
we assume the shoulder is sitting on top
of an approximately linear slope of the \Th peak:
take \Tl as the temperature where the curvature
of $c_V(T)$ is maximally negative (i.e.,
the local minimum of the curvature around the shoulder).
Note the sparse experimental data has been interpolated
by $4$th to $6$th order polynomial fittings.
}
\label{Fig:Tl} 
\end{figure}

In Fig.~\ref{Fig:Tl} we provide the scaling of the lower
characteristic temperature $\Tl$ vs. length $L$ for various YC6
lattices. As $L$ increases, the lower peak/shoulder structure
only slightly shifts towards lower temperatures,
while it also becomes more
pronounced, suggesting stronger
120$^\circ$ correlations in the system
[in accordance with $S(K)$ data in Fig.~3(e)
in the main text].

In the inset of Fig.~\ref{Fig:Tl} we plot the estimated
position for $\Tl$ vs. $1/L$, and
also indicate the position of shoulder-like structure in
experimental curves \cite{Cui2018}. Note that there exists some
arbitrariness in determining $\Tl$ from peak/shoulder structure,
as reflected in the error bars.
Furthermore, in
order to reduce the finite-length effect in YC6 data, we take the
difference between YC$6\times12$ and YC$6\times9$ data
(divided by the $6\times3$ extra sites) as a YC6 ``bulk" results,
whose lower peak is even more pronounced and the corresponding
$\Tl$ well agrees with experiments, up to error bars.

\begin{figure}
\includegraphics[angle=0,width=.45\linewidth]{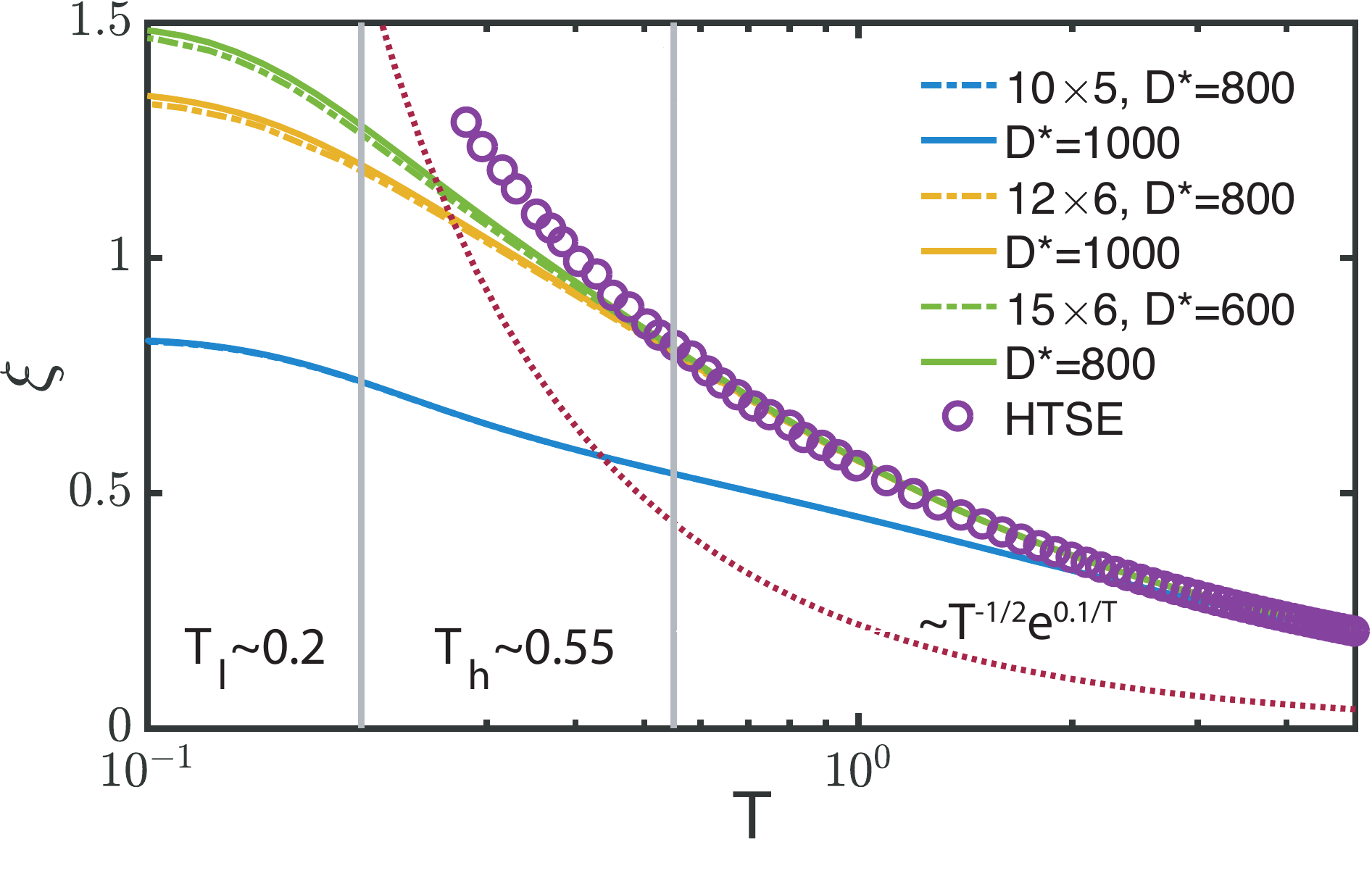}
\caption{
Correlation length, $\xi$, on various YC cylinders,
together with HTSE-2D data
\cite{Elstner-1993} for comparison.
For reference, we also added the exponential increase
$\xi\propto T^{-1/2} e^{\Tl/2T}$ predicted by
field-theoretical arguments (red dotted line)
\cite{Chakravarty1988}. 
}
\label{Fig:Xi} 
\end{figure}

\subsection{Correlation length vs. temperatures}
Assuming the Ornstein-Zernicke form of the static structure factor
$\Sq = S(q_0)/[1+\xi^2(q-q_0)^2]$, in the close vicinity
of the ordering momentum $q_0$, this defines the
correlation length
\cite{Elstner-1993,Elstner1994} as
\begin{equation}
\xi^2 \cong \left. \tfrac{1}{2S(q)} \tfrac{\partial^2 S(q)}{\partial q^2} \right|_{q=q_0}
= \tfrac{c^2_{q_0}}{2S(q_0)} \sum_j
  r_{0j}^2 \, e^{-i q_0 \cdot r_{0j}} \,
  \langle \bm{S}_0 \cdot \bm{S}_j \rangle
 \text{ ,}\label{Eq:Xi}
\end{equation}
where $r_{0j} \equiv r_j - r_0$ with $r_j$ the lattice location
of site $j$, with $j$ running over the whole lattice,
and again $r_0$ fixed in the center of the system.
The constant $c^2_{q_0} \equiv \langle \cos^2 \alpha_{0j}\rangle
\in [0,1]$ accounts for an angular
average with $\alpha_{0j}$ the angle in between
$q_0$ and $r_{0j}$.
In the present context of the TLH, we chose $q_0=K$
as the ordering momentum which leads to $c^2_0 = 1/2$.

From Fig.~\ref{Fig:Xi}, one observes that the correlation length
$\xi$ remains very short, in that it is below one
lattice spacing down to temperatures even below 
$\Th\sim0.55$ also for the wider systems,
in agreement with HTSE-2D in the thermodynamic limit.
At lower temperatures, the correlation length gets 
enhanced with increasing width $W$ and length $L$.
Given the very short correlation length of just a very few lattice
spacings for $\Tl<T<\Th$,
incipient order can be ruled out
in this intermediate regime.
Instead, we associate this regime with activated
RLEs (with minima at $M$), 
which suppresses the long-range 
order formation at $K$ and thus
leads to a short $\xi$.
As temperature is lowered further, the
correlation length $\xi$ is expected to
increase exponentially.
As for our XTRG data, $\xi$ keeps increasing down to
$T\sim0.1$, where it saturates
due to the finite system size. 

\subsection{Entanglement spectra vs. temperatures}

\begin{figure}[!tbp]
\includegraphics[angle=0,width=0.55\linewidth]{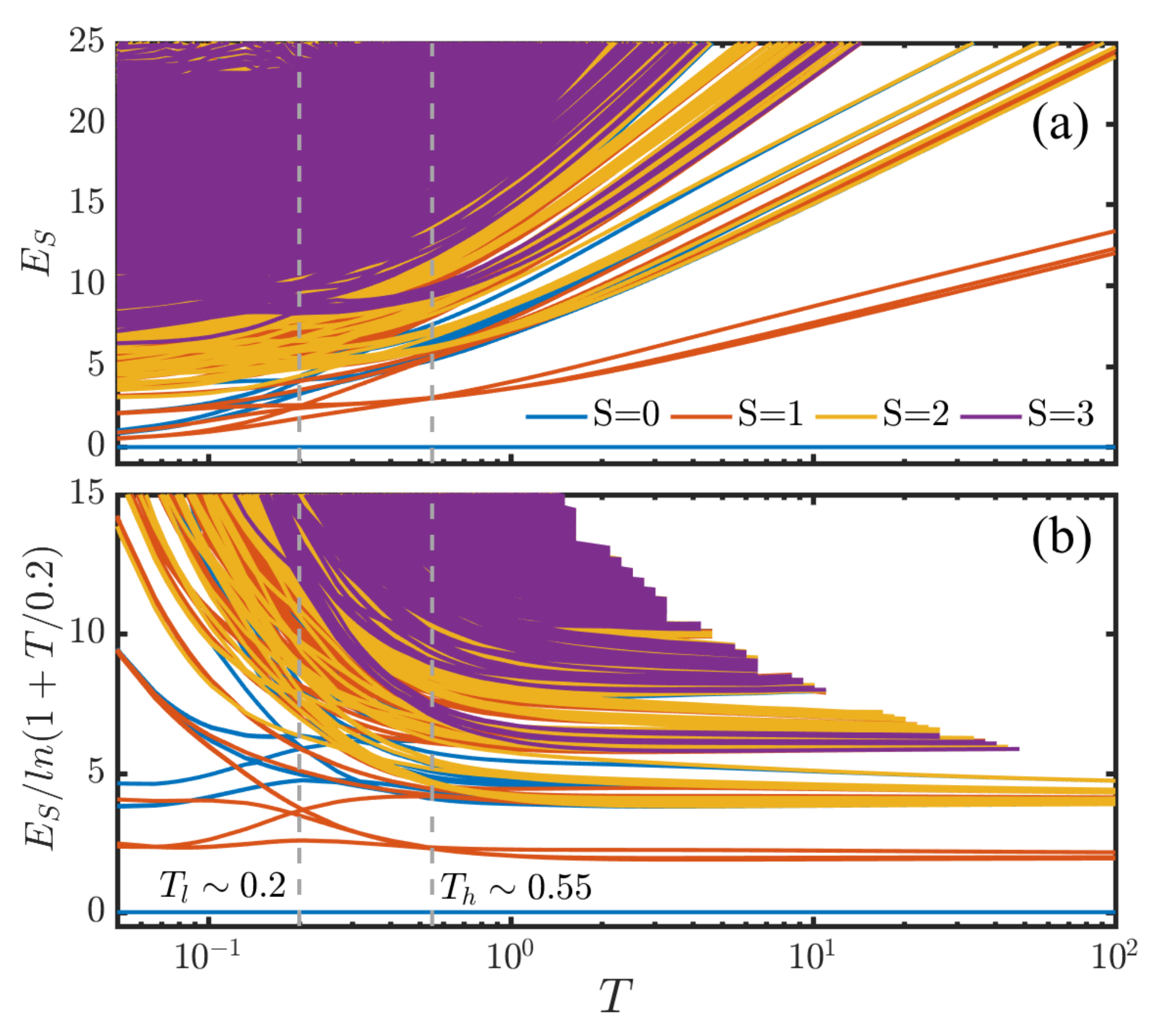}
\caption{
(Color online)
(a) Entanglement spectrum, $E_S$,
across a vertical cut
in the middle of a YC$6{\times}12$
system versus temperature $T$, where
the color differentiates the SU(2) spin symmetry
sectors as indicated in the legend.
Two temperature scales $T_l \sim 0.2$ and $T_h\sim0.55$
are indicated with the grey dashed lines.
(b) The same $E_S$ data, yet scaled 
by $\ln{(1+T/a)}$ with $a\sim \Tl $.
With this the high temperature spectra in (a) 
with $E_S \sim \ln T$ \cite{Chen2018} 
become horizontal lines.
At low $T<\Tl$, empirically, the chosen scale factor 
$1/\ln(1+\tfrac{T}{\Tl}) \sim 1/T$ still
represents a sensible scaling for the lowest
levels, thus suggesting $E_S\sim T$ there.
}
\label{Fig:RGFlow}
\end{figure}

\begin{figure}[!tbp]
\includegraphics[angle=0,width=0.85\linewidth]{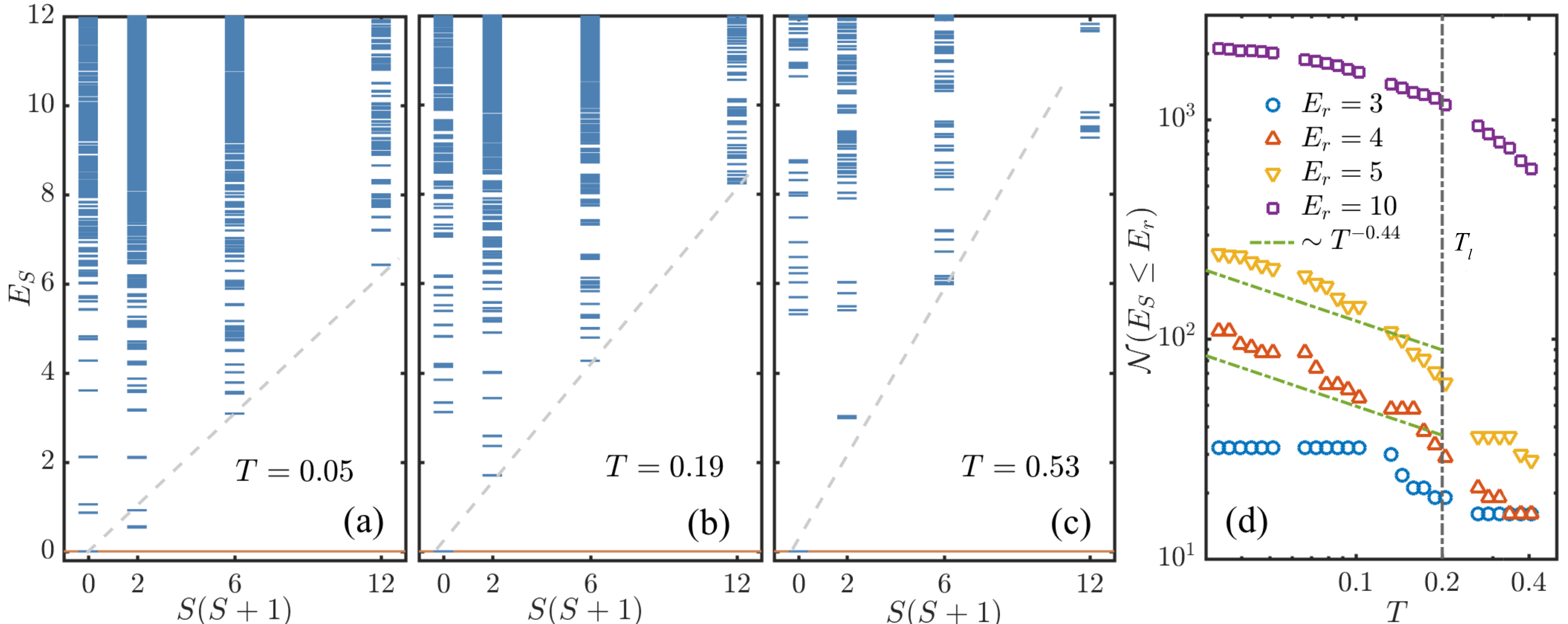}
\caption{
(Color online)
Entanglement spectrum ($E_S$)
of entanglement Hamiltonian
$\mathcal{H}_\mathrm{ES}$ with SU(2) spin
quantum number $S$ at
(a) $T=0.05$, (b) $T=0.19 \simeq T_l $,
and (c) $T=0.53 \simeq \Th$.
The dashed lines are guide for eyes,
indicating the increasing
slopes of the ``towers'' with increasing
temperatures.
(c) Counting the number of states
$\mathcal{N}(E_S\leq E_r)$
for $E_r = 3, 4, 5, 10$, at various temperatures.
It is observed that $\mathcal{N}$
increases as $T$ decreases,
and for a range of intermediate energies,
say, $E_r = 4, 5$, $\mathcal{N}$ scales
approximately polynomially vs. $T$ for $T<T_l$,
with $\sim T^{-0.44}$ {(see text)} shown as guid to the eye.}
\label{Fig:TOS}
\end{figure}

In Fig.~\ref{Fig:RGFlow},
we analyze the ``renormalization-group'' flow
of the MPO entanglement spectra
of the thermal state, $\rho(\beta)$ vs. the logarithm of
the energy scale $T$, for a vertical cut across the
middle of a YC$6{\times}12$ system,
with focus on the two temperature scales \Tl and \Th.
This is significantly more detailed than the single
number in terms of the MPO entanglement entropy $S_E$,
e.g., as analyzed in Fig.~3(f) in the main paper.
The entanglement spectra are also derived from the normalized
`purified' thermal state $|\rho(\beta)\rangle$
and its reduced density matrix $\mathcal{R}(\beta)_{I,I'}
\equiv \sum_{J} |\rho_{I,J}\rangle\langle\rho_{I',J}|$
\cite{Chen2018}, where $I,J$ represent degrees
of freedom in left and right
half of the system, respectively,
and $\beta$ is the inverse temperature.
Then, the entanglement spectrum, $\tilde{E}_S$,
at a given inverse temperature, $\beta$, is 
obtained by diagonalizing 
$\mathcal{H}_\mathrm{ES} \equiv -\ln{\mathcal{R}}$.
These are analyzed relative to their `ground-state
energy', i.e., $E_S \equiv \tilde{E}_S-E_0$.

In the entanglement spectra $E_S$ in \Fig{Fig:RGFlow},
the different symmetry sectors of the MPO
virtual bond states are differentiated
by color as indicated in the legend.
We show eigenstates for each
spin symmetry sector, up to a largest `energy' 
$E_S \leq 25$, which corresponds to a weight
in the density matrix as low as
$\ge e^{-25} \sim 10^{-11}$.
From Fig.~\ref{Fig:RGFlow},
we can see that, 
while in the high temperature regime
the levels in $E_S$ are rather far apart
(i.e. the MPO is close to a product state),
they become much more
dense with decreasing temperature.
They show systematic qualitative changes, and in particular
line-crossings in the `low-energy sector' around
$T_l$ and $T_h$. 
For $T<\Tl$, the entanglement spectra show
a systematic (algebraic) approach towards the `ground state',
with \Fig{Fig:RGFlow}(b) roughly suggesting $E_S\sim T$
for the lowest levels once $T<\Tl$.
These distinct behaviors of $E_S$
in different temperature regimes
are consistent with the existence
of two temperature scales.

In Fig.~\ref{Fig:TOS}(a-b), we study the spectroscopy of the
entanglement Hamiltonian $\mathcal{H}_\mathrm{ES}$.
The entanglement levels are plotted
with respect to different symmetry labels
from $S=0$ up to $S=4$,
at $T=0.19$ and $0.05$, respectively.
As indicated by the gray dashed lines,
the lowest level in each symmetry sector
decreases, roughly algebraically
as we cool down the system,
while also the slope systematically decreases.
As a consequence, the number of states with weights
$E_S < E_r$, denoted by $\mathcal{N}(E_S\leq E_r)$,
increases as $T$ decreases.
We plot $\mathcal{N}(E_S\leq E_r)$ in Fig.~\ref{Fig:TOS}(d)
with $E_r = 3, 4, 5, 10$. From this we observe that
$\mathcal{N}(E_S\leq E_r)$
increases roughly polynomially
when $T\leq T_l$,
indicated by the green dashed lines.
If furthermore, we assume rather heuristically
that for intermediate
$E_r$, say, $E_r=4$ or 5,
each level in the range $E_S\leq E_r$ contributes
crudely 
equally to
the thermal entropy $S_E$
of the entanglement Hamiltonian $\mathcal{H}_{ES}$ system,
this is 
consistent with an expected log-scaling
of the entanglement entropy,
$S_E \sim \ln{\mathcal{N}}
\propto a_{\mathcal{N}} \ln{\beta} +b$.
And, indeed, this hand-waving argument agrees with
the scaling $S_E \propto a\ln{\beta}+b$ 
for $T\leq T_l$, as observed and discussed with
Fig.~3(f) in the main text, 
even with roughly consistent 
slope $a_{\mathcal{N}} \approx a=0.44$, 
also shown in Fig.~\ref{Fig:TOS}(d) as guide to the eye.

\subsection{Spin wave analysis}
\begin{figure}[!tbp]
\includegraphics[angle=0,width=0.52\linewidth]{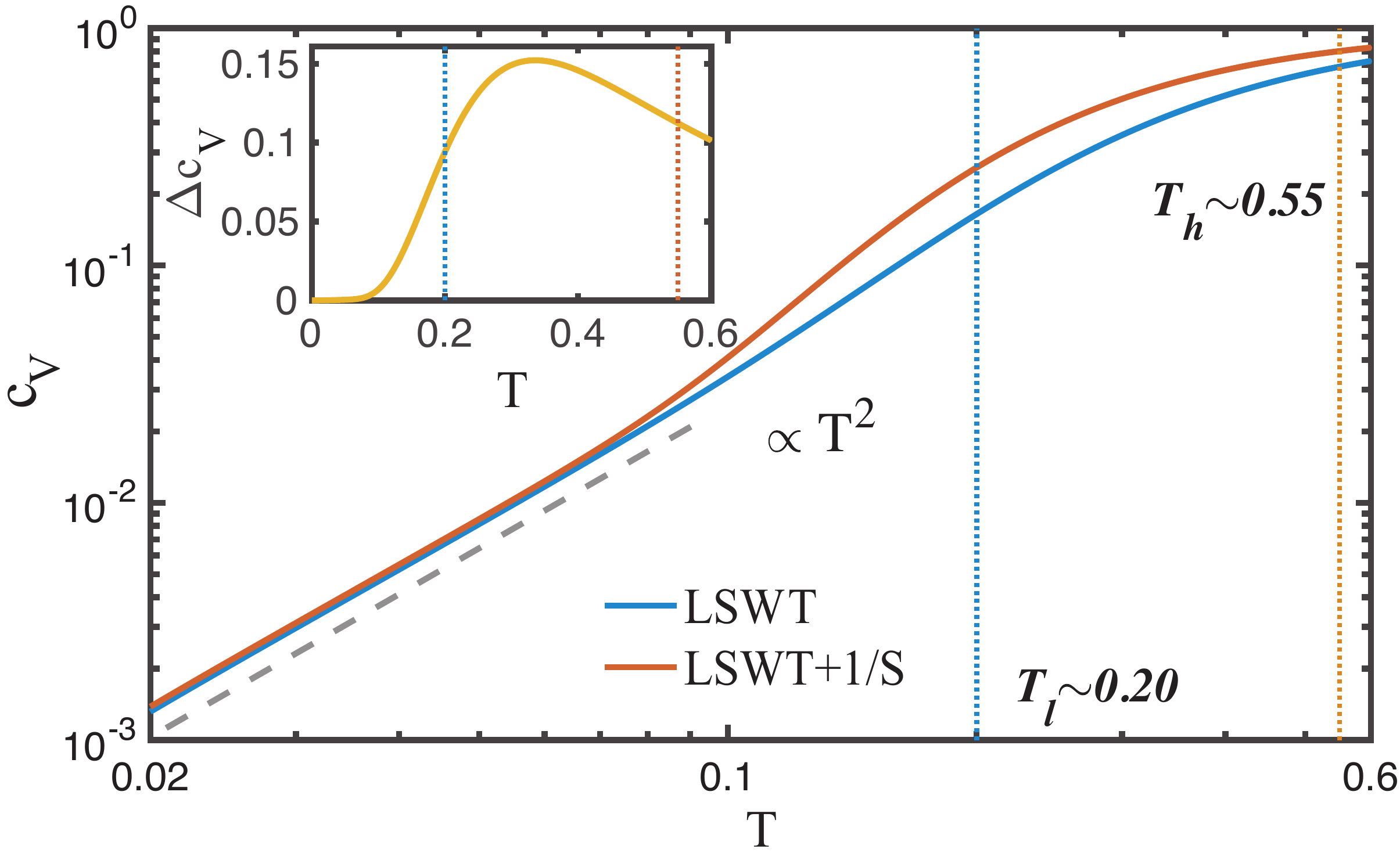}
\caption{(Color online) 
Low-temperature specific heat $c_V$
from the spin wave analysis, with and without $1/S$
corrections \cite{Starykh2006},
where the gray dashed line indicates the 
$T^2$  behaviors of $c_V$ at low $T$ due to the 
linear spin wave dispersion $E_k \propto k$ for small $k$. 
Inset shows the difference
between the two approximations. The vertical markers
indicate the two energy scales \Tl and \Th as specified.
}
\label{Fig:LSWT}
\end{figure}
Linear spin wave theory (LSWT) can very well 
capture the $120^{\circ}$ order in the ground state of TLH \cite{Jolicoeur-1989,Chubukov1994,Chernyshev-2009}.
It is thus also believed that LSWT is able to describe 
thermodynamics at very low temperatures. 
Here we take the spin wave spectra
with and without $1/S$ corrections
[Eq.~(12) from \cite{Starykh2006}], 
compute the specific heat according to 
the conventional Bose-Einstein 
distribution of magnon gas, and draw a comparison
to our two-temperature-scale scenario.

In Fig.~\ref{Fig:LSWT}, we can see that the
$c_V$ curve with $1/S$ corrections also exhibit a
``shoulder-like" structure  at around $T_l$,
below which it gradually changes into a $T^2$ scaling, 
coinciding there with pure LSWT results at low $T$. 

The $1/S$ correction gives rise to the 
difference $\Delta c_V$ which,
as manifested in the inset of Fig.~\ref{Fig:LSWT}, 
strongly affects the
intermediate temperature regime in
between $T_l$ and $T_h$.  
$\Delta c_V$ starts to decrease rapidly 
below $T\sim T_l$, and the influences of RLEs
as well as other renormalization effects in the spectrum 
due to $1/S$ corrections are largely absent below $T=0.1$.
This is consistent with our spin structure factor
as discussed with Fig.~3(c-e), where 
$S_M$, representing the activation of RLEs,
also becomes clearly weakened below $T_l$.

The LSWT analysis of $c_V$ and its $1/S$ correction 
in \Fig{Fig:LSWT} thus further 
confirms the existence of two temperature scales, 
and, in particular, the lower one $T_l$. 
Overall, \Fig{Fig:LSWT} provides useful 
complementary thermal data to our finite-size XTRG results.

\subsection{Chiral Correlations on YC4}

\begin{figure}[!tbp]
\includegraphics[angle=0,width=0.55\linewidth]{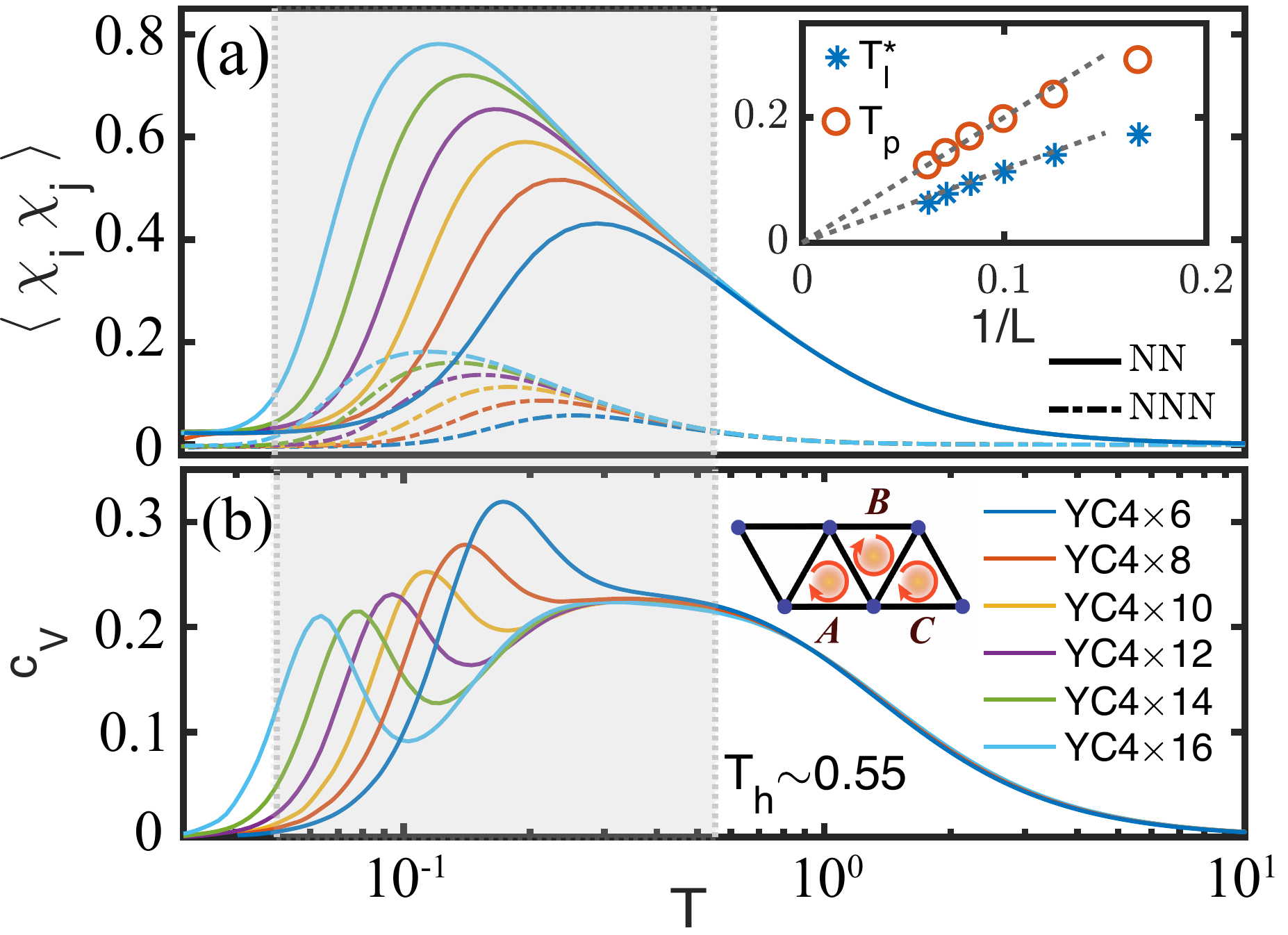}
\caption{(Color online)
(a) Chiral correlations compared to
(b) specific heat $c_V$
in YC4 for various lengths $L$.
In (a) we consider correlations 
between nearest- (NN) and next-nearest neighboring 
(NNN) triangles in the system center 
[i.e., using $i=A$ and $j=B,C$ 
as shown in the inset of (b) with the triangles
taken in the system center [cf. Fig.~1].
The disappearance
of the low-energy regime for $1/L\to0$
is specific to YC4 \cite{Chen2018}.
The inset shows finite size scaling of
the peak position, $T_p$,
in the chiral correlations $\langle \chi_i \chi_j \rangle$,
as well as of the temperature scale, $T_l^*$,
derived from the lower-temperature peak in $c_V$ from (b).
}
\label{Fig:ChiralYC4}
\end{figure}

This section complements the analysis of chiral correlations
in Fig.~5 in the main paper, in that we focus on the
case of YC4 which is special, in that $\Tl^\ast$ scales to zero
as $L\to\infty$ (hence the asterisk with \Tl).
From \Fig{Fig:ChiralYC4} we observe that short-ranged (NN and NNN) 
chiral correlations build up and become strong
at intermediate temperatures,
$\Tl^* \lesssim T \lesssim \Th$. 
Clearly, the peak at $T_p$ in the chiral correlations
[\Fig{Fig:ChiralYC4}(a)] correlates with the 
low-energy scale $\Tl^*$
derived from $c_V$ [\Fig{Fig:ChiralYC4}(b)]. 
Below $\Tl^*$, chiral correlations 
become negligibly small, 
which is also directly confirmed by
DMRG simulations (not shown).
While YC4 also enters an anomalous liquid-like
regime for $T<T_h$,
it becomes a stripe phase below a low-energy scale $\Tl^*$
which diminishes to zero
in the thermodynamic limit 
as demonstrated in the inset
to \Fig{Fig:ChiralYC4}(a) \cite{Chen2018}.
This is qualitatively different
from $\Tl \sim 0.20$ in wider cylinders and strips, 
where it becomes a stable temperature scale vs.
various system sizes and boundary conditions
as shown and discussed with Fig.~2(a).
Importantly, \Fig{Fig:ChiralYC4} demonstrates that
the peak $T_p$ in the chiral correlations 
closely follows the low-energy
scale \Tl derived from the specific heat. In this sense
we conclude that they can be ascribed to the same
crossover scale that separates the low-energy
(here stripe) order from the intermediate regime and,
moreover, that finite chiral correlations constitute
a characteristic property of the intermediate regime.

\begin{figure}[!tbp]
\includegraphics[angle=0,width=0.52\linewidth]{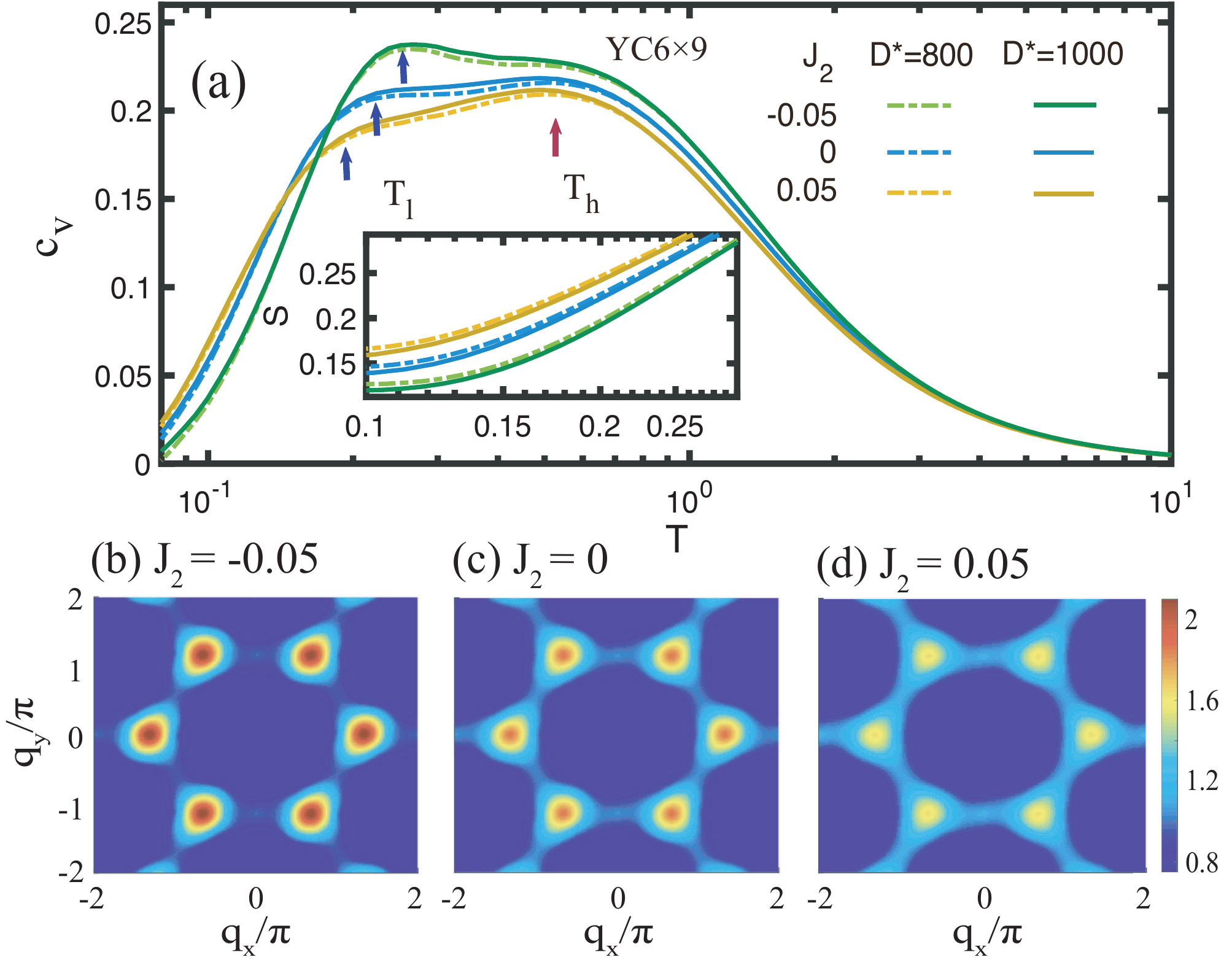}
\caption{(Color online)
(a) Specific heat, as well as thermal entropy (inset), 
of $J_1$-$J_2$ TLH, with $J_2= -0.05$, $0$, 
and $0.05$, on YC$6\times 9$ geometry.
(b,c,d) show the structure factors at
$T \simeq0.53$ near the high temperature scale $\Th$.
}
\label{Fig:TriJ2Cm}
\end{figure}

\section{Tune spin frustration by deforming the triangular-lattice Heisenberg model}
\subsection{$J_1$-$J_2$ Triangular lattice Heisenberg model}

To shed further light on
the two-temperature-scale scenario,
we deform the TLH by including
a small but finite next-nearest coupling $J_2$,
\begin{equation}
H = J_1 \sum_{\langle i, j \rangle}  \vec{S}_i \vec{S}_j  + J_2 \sum_{\langle\langle i,j\rangle\rangle}  \vec{S}_i \vec{S}_j,
\end{equation}
having $J_1 \equiv J=1$.
A small AF coupling $J_2>0$
adds frustration, and hence suppresses
120$^\circ$ ordering
on a three-sublattice configuration
\cite{Zhu2015,Hu2015,Iqbal2016,Gong2017},
while $J_2<0$ reinforces
the ferromagnetic
correlations between two
spins on the same $120^\circ$ sublattice. 

In \Fig{Fig:TriJ2Cm} we compare
the specific heat $c_V$ of YC$6\times9$,
for $J_2=0$ with $J_2=\pm 0.05$.
From the temperature dependence in \Fig{Fig:TriJ2Cm}(a)
we can see that as we change $J_2$
from $-0.05$ to $0.05$,
the $\Tl$ peak systematically shifts
to lower temperatures, while at the same time, it weakens.
In contrast, the broad peak at $T_h$
keeps its position while varying $J_2$. The overall
downward shift for temperatures $T\gtrsim 0.2$
and upward shift for very small temperatures
suggests that with increasing $J_2$,
more entropy is
transferred to lower temperatures.
Indeed, as shown in the
inset of \Fig{Fig:TriJ2Cm}(a),
compared to $J_2=0$ case,
thermal entropy $S$ at $T=0.1$ is
increased (decreased) by $\sim
15\%$, when $J_2=0.05$ ($-0.05$) is introduced.

The sensitivity of the structure factor on
$J_2$ is analyzed in \Figs{Fig:TriJ2Cm}(b-d).
As expected, increasing $J_2$ significantly
reduces the weight $S(K)$, but enhances the weight $S(M)$.
This is consistent with dynamical calculations of the
$J_1$-$J_2$ TLH \cite{Ferrari2019},
as well as that includes spin XXZ anisotropy
\cite{Ghioldi2015}, which show 
that RLEs are  
renormalized downward by increasing $J_2>0$, 
giving rise to an extended dispersive
continuum above the ``roton" minimum. 

\begin{figure}[!tbp]
\includegraphics[angle=0,width=0.52\linewidth]{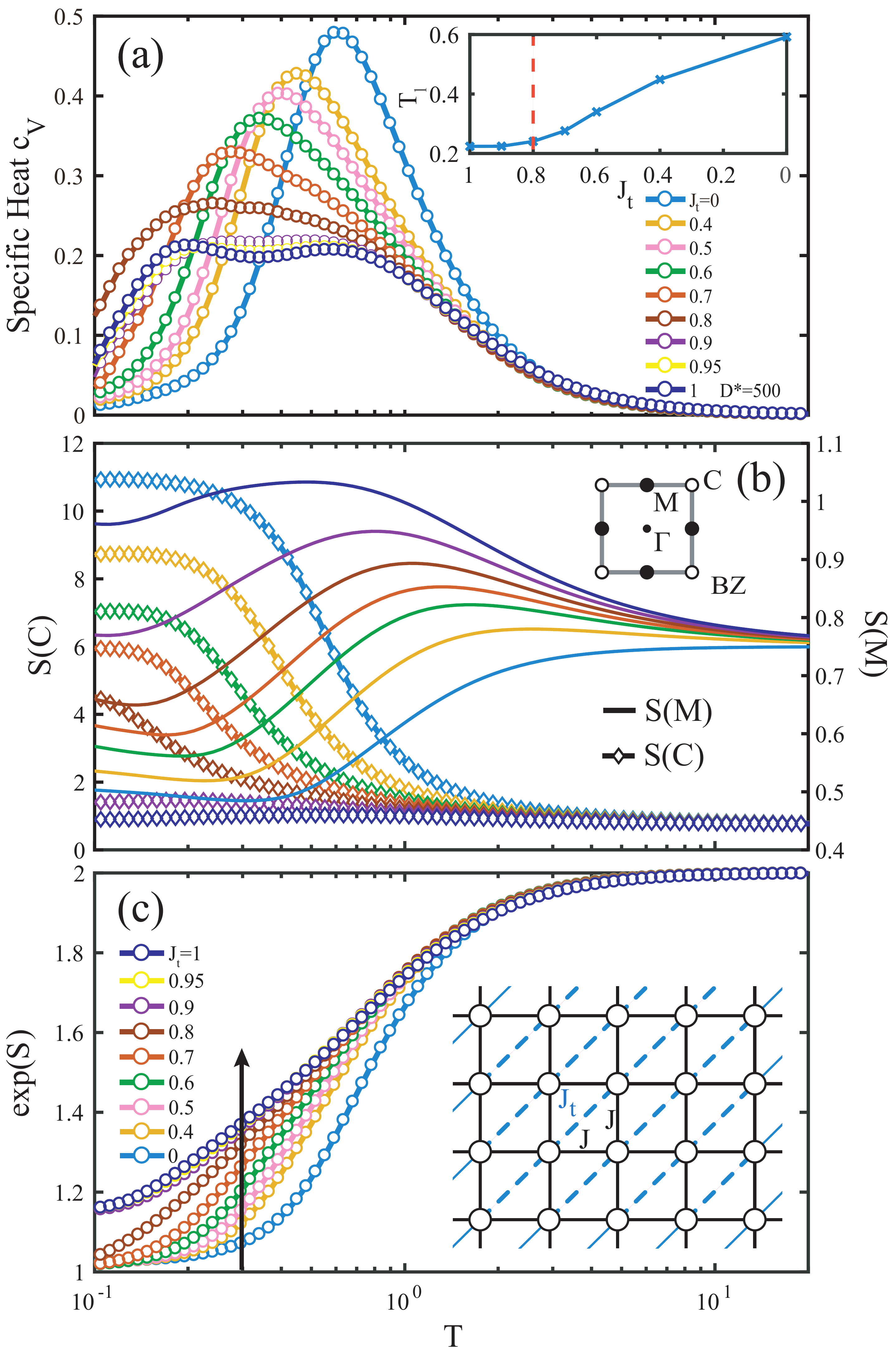}
\caption{
(Color online
(a) Specific heat data for the frustrated SLH on a $6 \times 12$
    cylinder with $J_t$ varied from 0 to 1,
    keeping up to $\Dstar = 500$
    multiplets [$1790$-$2010$ U(1) states].
    The inset traces the
    low-temperature peak (or shoulder)
    in the $c_V$ curves.
(b) Static structure factor $S(q)$ at $C = (\pi, \pi)$
    (lines with markers) 
    and $M = (0, \pi)$ (solid lines). 
    The constant offset towards large $T$ is a trivial finite size
    effect: it is the same for $S(M)$
    and $S(C)$ (note the significantly different vertical scales
    for $S(C)$ [left axis] and $S(M)$ [right axis])
    and comes from the $r_j=r_0$ contribution
    in the definition of the structure factor, which gives
    a constant $\langle \bm{S}^2 \rangle_T =\tfrac{3}{4}$.
(c) Thermal entropy $S$ decreases 
    from $\ln 2$ as $T$ is lowered. The thermal entropy
    moves to larger values as the frustration is
    turned on via $J_t$.
}
\label{Fig:FSLH}
\end{figure}

\subsection{Frustrated square-lattice antiferromagnet}
\label{Sec:FSLH}

Besides adding $J_2$ in to TLH,
we can also deform the Hamiltonian towards
the square lattice Heisenberg (SLH)
model via a tuning parameter $J_t$
where $J_t=1$ corresponds
to the TLH, and $J_t=0$ to the SLH.
The corresponding Hamiltonian of
the frustrated SLH is given by,
\begin{equation}
H = J \sum_{\langle i,j \rangle} \vec{S}_i \cdot \vec{S}_j + J_t \sum_{\langle\langle i,j \rangle\rangle} \vec{S}_i \cdot \vec{S}_j,
\label{Eq:SLH}
\end{equation}
where $\langle .,. \rangle$
denotes nearest-neighbor,
and $\langle\langle .,. \rangle\rangle$
next-nearest
neighboring pairs of sites along
one of the diagonals only
[e.g., as depicted in the inset to
 \Fig{Fig:FSLH}(c)].
Their respective coupling strengths
are given by $J$ and $J_t$,
where again $J=1$ sets the unit of energy,
unless specified otherwise.
The system features square lattice N\'eel
order at $T=0$ for $J_t=0$,
leading to low-$T$ RC behavior \cite{Manousakis1991}.
Finite $J_t>0$ introduces
frustration in the system,
which leads to an abrupt
onset of an incommensurate
spiral wave at some critical
$\Jtc\sim 0.79$ \cite{Wb11incomm}.
For the isotropic TLH at $J_t=1$
this turns commensurate
and yields the familiar $120^\circ$ order.
Our results for the SLH on a
$6\times12$ cylinder
(i.e. $L=2W=12$) are summarized
in \Fig{Fig:FSLH},
where we vary $J_t$ from $0$ to $1$
(see legend), with $\Dstar = 500$
multiplets kept.

The specific heat $c_V$ 
is shown in \Fig{Fig:FSLH}(a).
The tracked lowest temperature scale,
i.e., global peak position for $J_t\leq0.8$
and the position of the lower shoulder
for $J_t > 0.8$,
is plotted as `\Tl' in the inset.
This scale starts from about $\Tl = \Th\sim0.6$ 
for $J_t = 0$ (i.e., the pure SLH case).
While $\Tl$ decreases gradually with increasing $J_t$,
at the same time a shoulder emerges at the original
peak position \Th which only marginally moves to smaller values.
At around $J_t\simeq0.8$, the single peak in the specific
heat for small $J_t$ is about to fully split into two peaks.
There \Tl has already nearly also
reached its final value of $0.2$
for $J_t = 1$, i.e., the isotropic TLH.
Interestingly, the critical value for
which ground state calculations observe the onset of
incommensurate correlations, $\Jtc\sim0.79$ \cite{Wb11incomm},
roughly coincides with the $J_t$ for which a well separated
two-peak structure has developed at finite $T$
(for the finite size systems here even with a minimum
in between), where \Tl is already also close to its final
value of 0.2 for the TLH.

To understand the physical meaning of the temperature scales and their behaviors under various $J_t$, we look at 
the static structure factors, $S(C)$ and $S(M)$, in \Fig{Fig:FSLH}(b).
The points $C$ and $M$ in reciprocal space
are pointed out in the inset. 
We observe that, for $J_t \lesssim \Jtc$,
the AF magnetic order [at $C = (\pi, \pi)$, i.e., N\'eel order] melts
most rapidly at the characteristic temperature $\Tl$.
The magnetization per site may also be estimated
in the present case by $m \approx \sqrt{S(C)/N}$
(note the slightly different normalization due to angular
average as compared to the TLH).
For $J_t=0$, this yields $m \sim 0.39$
which overestimates the thermodynamic limit $m\sim0.31$
\cite{White07} due to finite-size effects.

The stable large energy scale at \Th relates to different physics,
here argued to be RLEs.
This argument can be solidified by analyzing 
the structure factor at the point 
$M = (0,\pi)$ which, indeed, shows anomalous enhancement
at intermediate temperatures for finite $J_t$. 
Note that the position of the maximum in 
$S(M)$ changes from $\sim 1.5$ for $J_t = 0.6$
to the value of $\Th \sim 0.55$ itself for $J_t = 1$.

In \Fig{Fig:FSLH}(c), finally, 
we still present data for the thermal 
entropy $S$ for various $J_t$ values.
From this we can see an anomalous enhancement of the 
entropy 
as $J_t \gtrsim \Jtc\sim 0.8$. 
It is remarkable to find the zero temperature
commensurate-incommensurate transition also reflected
here as a (residual) entropy enhancement in our
thermal calculations at our lowest temperatures.

In summary, the specific heat 
as well as the static structure factor data in
\Fig{Fig:FSLH} provide further strong support
for the emergence of a two-temperature-scale scenario
at sufficiently large frustration $J_t$, 
including the $J_t=1$ TLH.

\end{document}